\documentclass[manuscript]{acmart}

\usepackage{comment}
\usepackage{booktabs}
\usepackage{graphicx}
\usepackage{listings}
\usepackage{threeparttable}
\usepackage{xcolor}
\usepackage{array}
\usepackage{ragged2e}
\usepackage{hhline}
\usepackage{colortbl}
\usepackage{svg}
\usepackage{amsmath}
\usepackage{hyperref}
\usepackage{microtype}
\usepackage{enumitem}

\AtBeginDocument{%
  \providecommand\BibTeX{{%
    \normalfont B\kern-0.5em{\scshape i\kern-0.25em b}\kern-0.8em\TeX}}}

\usepackage{pdflscape}%sideways picture
\usepackage{svg}%import svg
\usepackage{amsmath} %import svg
\usepackage{array}
\usepackage{arydshln}
\colorlet{tableheadcolor}{gray!25} % Table header colour = 25% gray
\colorlet{tablerowcolor}{gray!5} % Table row separator colour = 10% gray
\colorlet{tablerowcolor2}{gray!12} % Table row separator colour = 10% gray
\colorlet{tablerowcolor3}{gray!25} % Table row separator colour = 
\newcommand{\rowcollight}{\rowcolor{tablerowcolor2}}
\definecolor{grey}{HTML}{969696}
\definecolor{dgrey}{HTML}{01665e}
\definecolor{lgrey}{HTML}{5ab4ac}
\definecolor{improveCol}{HTML}{253494}
\definecolor{worsenCol}{HTML}{d7191c}
\definecolor{hiCol}{HTML}{d01c8b}
\definecolor{loCol}{HTML}{4dac26}
\definecolor{lgreen}{HTML}{f0f9e8}
\definecolor{dgreen}{HTML}{005a32}
\AtBeginDocument{%
  \providecommand\BibTeX{{%
    \normalfont B\kern-0.5em{\scshape i\kern-0.25em b}\kern-0.8em\TeX}}}

\renewcommand\footnotetextcopyrightpermission[1]{} % removes footnote with conference information in first column
\begin{document}

\title{Seamful XAI: Operationalizing Seamful Design in Explainable AI}

\author{Upol Ehsan}
\affiliation{%
  \institution{Georgia Institute of Technology}
  \city{Atlanta}
  \state{GA}
  \country{USA}}
 % \email{ehsanu@gatech.edu}
 
\author{Q. Vera Liao}
\affiliation{%
  \institution{Microsoft Research}
  \city{Montreal}
  \state{}
  \country{Canada}}
 % \email{philipp.wintersberger@carissma.eu}

\author{Samir Passi}
\affiliation{%
  \institution{Microsoft}
%   \city{Princeton}
  \state{}
  \country{USA}}
 \email{}

 \author{Mark O. Riedl}
\affiliation{%
  \institution{Georgia Institute of Technology}
  \city{Atlanta}
  \state{GA}
  \country{USA}}
 \email{}

 \author{Hal Daumé III}
\affiliation{%
  \institution{University of Maryland \& Microsoft Research}
  \city{}
  \state{}
  \country{USA}}
 \email{}


\begin{abstract}
Mistakes in AI systems are inevitable, arising from both technical limitations and sociotechnical gaps. While black-boxing AI systems can make the user experience seam\textit{less}, hiding the \textit{seams} risks disempowering users to mitigate fallouts from AI mistakes. Instead of hiding these AI imperfections, can we leverage them to help the user? While Explainable AI (XAI) has predominantly tackled algorithmic opaqueness, we propose that seamful design can foster AI explainability by revealing and leveraging sociotechnical and infrastructural mismatches. We introduce the concept of \textit{Seamful XAI} by (1) conceptually transferring “seams” to the AI context and (2) developing a design process that helps stakeholders anticipate and design with seams. We explore this process with 43 AI practitioners and real end-users, using a scenario-based co-design activity informed by real-world use cases. We found that the Seamful XAI design process helped users foresee AI harms, identify underlying reasons (seams), locate them in the AI's lifecycle, learn how to leverage seamful information to improve XAI and user agency. We share empirical insights, implications, and reflections on how this process can help practitioners anticipate and craft seams in AI, how seamfulness can improve explainability, empower end-users, and facilitate Responsible AI. 
\end{abstract}

\begin{CCSXML}
<ccs2012>
   <concept>
       <concept_id>10003120.10003123.10010860.10010883</concept_id>
       <concept_desc>Human-centered computing~Scenario-based design</concept_desc>
       <concept_significance>300</concept_significance>
       </concept>
   <concept>
       <concept_id>10003120.10003121.10011748</concept_id>
       <concept_desc>Human-centered computing~Empirical studies in HCI</concept_desc>
       <concept_significance>500</concept_significance>
       </concept>
   <concept>
       <concept_id>10003120.10003121.10003126</concept_id>
       <concept_desc>Human-centered computing~HCI theory, concepts and models</concept_desc>
       <concept_significance>500</concept_significance>
       </concept>
   <concept>
       <concept_id>10003120.10003130.10003131</concept_id>
       <concept_desc>Human-centered computing~Collaborative and social computing theory, concepts and paradigms</concept_desc>
       <concept_significance>300</concept_significance>
       </concept>
   <concept>
       <concept_id>10010147.10010178</concept_id>
       <concept_desc>Computing methodologies~Artificial intelligence</concept_desc>
       <concept_significance>300</concept_significance>
       </concept>
 </ccs2012>
\end{CCSXML}

\ccsdesc[300]{Human-centered computing~Scenario-based design}
\ccsdesc[500]{Human-centered computing~Empirical studies in HCI}
\ccsdesc[500]{Human-centered computing~HCI theory, concepts and models}
\ccsdesc[300]{Human-centered computing~Collaborative and social computing theory, concepts and paradigms}
\ccsdesc[300]{Computing methodologies~Artificial intelligence}
\keywords{Seamful design, explainable AI, responsible AI, AI ethics, human-AI interaction, explanations}


\maketitle

\section{Introduction}
\begin{quote}
   \textit{ ```We make it so easy to use that \textit{you won't even have to think about it!}' exclaimed the VP of a fin-tech company while proudly pitching the latest AI-powered tool. `When you use our product, it's seamless! One click, that's it. Plug and play! It's frictionless!'''---firsthand account of an informant in our study.}
\end{quote}

\noindent
Mistakes in AI systems are inevitable. Handling the mistakes is harder when the system’s decision-making is hidden or black-boxed. Seamless design---an often unquestioned design virtue \cite{Chalmers2004, inmanribes2019}---can be a double-edged sword. On the positive side, it promotes simplicity and ease of use~\cite{inmanribes2019, seamfulviz2022}. On the negative side, it abstracts and conceals important factors about the AI system that are crucial for explainability. These factors can be both inside AI's black box---such as its training and algorithms---or outside---such as sociotechnical infrastructures like data and integration practices. Breakdowns in AI systems often occur when the assumptions we make in design and development do not hold true when they are deployed in the real-world. For example, an AI system can fail when it’s trained on data from North America but deployed in South Asia, especially when the end user is unaware of this infrastructural mismatch. Concealing these mismatches through a seamless ideal can lead to downstream harms for end-users, such as unquestioned AI acceptance. What can we do differently? How do we move beyond seamless AI? And what do we gain by doing so?

One strategy would be to critically question the seamless ideal and, instead of hiding an AI system's seams, leverage them in technology design. Here, the concept of {\em seamful design}~\cite{Chalmers2003seamful, Chalmers2004}, a complement (not a rival) of seamless design~\cite{inmanribes2019}, can help. With conceptual roots in Ubiquitous Computing (Ubicomp), seamful design deals with seams---``mismatches’’, ``gaps’’, ``cracks’’, and ``uncertainties’’ that arise when a technology is deployed in the real world~\cite{BrollBenford2005, phoebeseamless2022}. Instead of hiding seams or treating them as problematic, seamful design argues for \textit{strategically revealing} (and concealing) seams to support user agency, re-configuration, and appropriation. For instance, a classic example is a "seamful map" revealing varying strength of WiFi coverage in a space. Knowing exactly where the Wi-Fi is weakest will allow a user to best use it. Thus, awareness of the blind spots and uncertainties in the technology can help users leverage the technology better. 

While there has been commendable work on seamful design in Ubicomp~\cite{Chalmers2003, Chalmers2003seamful, Chalmers2004, BrollBenford2005} and increased calls for it in AI~\cite{kaur2022sensible,ehsan2021explainability}, key questions remain unexplored: how can we define seams in AI? How may we find them before deployment? Where do they appear in the AI's lifecycle?  How may we use them to augment agency and explainability? We tackle these questions in this paper.

We introduce the notion of \textit{Seamful Explainable AI} by transferring seamful design into AI and exploring its practical intersection with the field of Explainable AI (XAI). We operationalize the notion of Seamful XAI by (1) conceptually transferring “seams” to the context of AI systems and (2) developing a design process that helps stakeholders identify and use the seams to enhance AI explainability and user agency. We chose to bridge seamful design and XAI because the two areas have a natural connection: both challenge the notion of black-boxing. In fact, much of XAI’s proverbial “opening of the black box” can be viewed as undoing the seamlessness of AI systems. 

In the context of AI, we define and conceptualize seams as \textit{mismatches and cracks between assumptions made when designing and developing the AI system and the reality of its deployment context}. The following example, inspired by real-world automated lending cases from our informants, illustrates how revealing seamful information can facilitate explainability and support user agency:

\begin{quote}
  \textit{The loan application for Ahmed, a great customer, is rejected by the AI-powered system. Nothing about the AI's feature-level explanation seems off to Nadia, the loan officer. Behind the scenes, there was a mismatch (seam): during the last system update in 2020, applicants could only have three active loans, but regulatory changes in 2021 allow up to five. Ahmed has four active loans, which triggered the denial. Had Nadia been aware of this mismatch (seam), she would be empowered to trigger a manual underwriting review.}

\end{quote}{}

\noindent
The type of understanding afforded by seamful information exceeds what is offered by algorithmic explanations. In the example, Nadia’s awareness of the AI’s blind spot, where model updates lag organizational policy shifts, can help her understand the ``\textit{why not}'' question---why the AI could not take certain things into consideration, instead of just the ``\textit{why}'' question answered by the feature-level explanation. This understanding equips Nadia to seek actions and contest the AI.

Our conceptualization of Seamful XAI is informed by and adds to two aspects: the {\em sociotechnical perspectives} of Human-centered XAI (HCXAI)~\cite{ehsan2020human,liao2021human,ehsan2021operationalizing} and the  {\em proactive or anticipatory} design stance in Responsible AI (RAI) practices~\cite{ehsan2022algorithmic, metcalfimpact2021, amy2022, Rakova2021, Madaio2020}. In a nutshell, we operationalize seams as a design element that could facilitate end-user explainability of the sociotechnical system---the human-AI assemblage~\cite{ehsan2021expanding}---rather than \textit{just} the technical system or algorithmic black-box. In contrast  to prior work in Ubicomp that addresses seams after they appear in deployment~\cite{BrollBenford2005,Chalmers2004,nilsson2016applying}, our focus is proactive and holistic (pre- and post-deployment). 
While the end goal of many RAI approaches is anticipating harms, the end goal of our process extends the current paradigm holistically: beyond anticipating harms (breakdowns), it also articulates the reasons (seams) that lead to breakdowns, traces where in the AI lifecycle these seams may appear, and most importantly, aims to \textit{turn a negative into a positive--using the seam in an opportunistic manner to augment user agency and explainability}. Stemming from the ethos of \textit{Jugaad} (Frugal) Innovation~\cite{radjou2012jugaad}, the aim is to be resourceful and strategically use the seams as part of the solution instead of trying to hide them. Thus, our work creates a new design space of ``seams'' and an actionable design process to facilitate XAI and RAI practices (elaborated further in Section~\ref{related-work}). 

Below we first situate the notion of Seamful XAI in the broader literature in seamful design, XAI, and RAI. Based on these perspectives, we introduce our proposed design process in Section~\ref{sec:sxai design process}. In Section~\ref{mehods}, we share our methods for conducting an empirical study to apply the design process. Finally, we share findings from the study with 43 users and practitioners of diverse roles. This paper presents a foundational step toward Seamful XAI. Our research contributions have 3 layers:

\begin{itemize}[nolistsep,noitemsep,left=0.5em]
    \item \textit{Conceptual:} 
    By introducing the notion of Seamful XAI, this paper breaks new ground and presents the pioneering effort to (a)~conceptually translate and (b)~explicitly define seams in the context of AI technologies--~uncharted areas despite calls for seamful design in AI~\cite{ehsan2021explainability, kaur2022sensible}.
    \item \textit{Methodological:} To operationalize Seamful XAI, we offer a design process that allows stakeholders to proactively identify seams and use them to augment explainability and user agency. 
    \item \textit{Empirical:} The findings and implications of our co-design study with 43 AI practitioners and end-users showcase that Seamful XAI aids participants in anticipating and designing with seams, enhancing explainability and user agency, promoting collaboration, and facilitating harm mitigation.
\end{itemize}

\section{Background and Related Work}
\label{related-work}
\subsection{Seamless and seamful design}
\label{seams-review}
Our understanding of seamless and seamful design follows research in the field of Ubiquitous Computing that explores how users live with heterogeneous technological systems~\cite{Chalmers2003, Chalmers2003seamful, Chalmers2004, BrollBenford2005,gaver2003ambiguity}. Seamlessness is almost always considered a good, even commendable, design ideal~\cite{inmanribes2019}. Technology builders strive to provide users with seamless experiences—a click and you get \textit{what} you want without having to think about \textit{how} and \textit{why}~\cite{inmanribes2019, seamfulviz2022}.
Technological systems, however, are complex sociotechnical assemblages. They do not exist in isolation, and intertwine with other components, practices, and systems in consequential ways. Seamless experiences thus require black-boxed systems shielding users from the intricate logic and messy machinery of technologies~\cite[p.~8]{phoebeseamless2022}. 

While seamless design emphasizes ease of use and concealment of messiness, seamful design prioritizes “configurability, user appropriation, and revelation of complexity, ambiguity, or inconsistency”~\cite[p.~1]{inmanribes2019}. Seams—mismatches, gaps, and uncertainties—are the inevitable ``cracks and bumps'' between heterogeneous components of socio-technical systems~\cite{BrollBenford2005, phoebeseamless2022}. Seamful design focuses on leveraging them in ways that help users make better decisions around technology use~\cite{Chalmers2003, inmanribes2019, Chalmers2003seamful, seamfulviz2022}. Central to seamful design in the notion of \textit{strategic revelations and concealment}~\cite{inmanribes2019,weiser1994creating}. It is not just about revealing seams but also about concealing them. Most importantly, we have to be strategic (intentional) about it. The goal is to support user agency, acknowledging that despite following the best processes, technology creators face uncertain downstream trajectories of use and must give users the resources to act effectively~\cite{inmanribes2019}. The question, however, isn’t just about \textit{what} seams to reveal (and what to hide), but also \textit{how} and \textit{when} to reveal them~\cite{inmanribes2019}.

Our work builds on and contributes to these different strands of scholarship by situating seamful design in the context of  AI. In doing so, we not only put forward a design process to anticipate, craft, and design with seams, but also make visible the values that seamful design provides for helping users work \textit{with} and \textit{through} the everyday failures and inevitable imperfections of AI systems.   

\subsection{Human-centered Explainable AI}

Our work also aims to augment AI explainability. The need to support user understanding through explainability is paramount given the opaqueness of AI technologies and their ability to cause harms at scale without human scrutiny. This pressing need gives rise to the field of XAI~\cite{gunning2019xai}, which has made commendable progress in producing a growing collection of techniques to enable algorithmic explanations. While the technical landscape of XAI is increasingly broad (for an overview of XAI techniques, see~\cite{guidotti2018survey,arrieta2020explainable, gilpin2018explaining,lipton2016mythos}), the majority of XAI research typically aims to address user questions such as ``how does the model make decisions'' or ``why does the model make a particular decision''~\cite{liao2020questioning} through means such as revealing how the model weighs different features and what rules it follows.

Despite its progress, XAI as a technical field has had its fair share of critique. Empirical studies examining the effectiveness of AI explanations in user interactions have found mixed results and even pitfalls of XAI. For examples, recent HCI studies repeatedly found that popular types of XAI techniques can risk creating over-trust and over-reliance when model predictions are wrong~\cite{zhang2020effect,poursabzi2021manipulating,bansal2021does, passi2022overreliance}. Technical explanations may in fact burden people who do not have the capability or motivation to engage~\cite{szymanski2021visual,ghai2021explainable} and risk impairing their user experience. These pitfalls have been attributed to the techno-centric focus of the current XAI field~\cite{liao2021human,ehsan2021explainability}, pursuing technical advancements without a clear understanding of and aiming to support the end-goals that people seek explanations for. In response to these limits, the area of HCXAI has emerged within the HCI community~\cite{ehsan2020human,liao2021human,ehsan2021operationalizing,wang2019designing}. These works center the development of XAI technologies around people's needs and end-goals, whether the end-goal is improving the model~\cite{narkar2021model}, making better decisions~\cite{xie2020chexplain,jacobs2021designing}, assessing model biases~\cite{dodge2019explaining}, seeking recourse~\cite{karimi2021algorithmic}, contesting the model~\cite{lyons2021conceptualising}, and so on~\cite{suresh2021beyond,liao2022connecting}. 

Inherent in HCXAI is also a sociotechnical perspective~\cite{ehsan2020human}: AI systems are often situated in socio-organizational environments, and people’s understanding must span factors both inside and outside the algorithmic black-box;  thus, explainability of the socio-technical elements---the human-AI assemblage~\cite{ehsan2021expanding}---needs more than just algorithmic transparency. This sociotechnical perspective has brought prior work to ``expand explainability''---defining new design spaces concerning the sociotechnical factors that govern the use of AI~\cite{liao2021human,ehsan2021operationalizing, BansalAccuracy, GonzalezExplanations}. For example, ~\citet{dhanorkar2021needs} highlight how explainability needs evolve and depends on who needs what and when.~\citet{ehsan2021expanding} challenge the algorithm-centrism and propose a design framework for social transparency of AI, by leveraging information about how other people interact and reason with the AI system.

In this work, we introduce seams as a new design space for HCXAI---facilitating end-user understanding of possible mismatches in and breakdowns of the sociotechnical system. In doing so, seams can complement existing explainability efforts by helping users understand not just why the AI did something, but also why the AI may not or cannot do something. Focusing on AI supporting human decision-making 
(vs. complete automation), we aim to support the end-goal of actionability for decision-making~\cite{liao2020questioning} and algorithmic contestability~\cite{lyons2021conceptualising}.

\subsection{Responsible AI in Practice}
Our work is also informed by, and contributes to,RAI practices. RAI is concerned with putting theoretical principles of AI ethics into practice~\cite{shneiderman2021responsible, Rakova2021, benjamins2019responsible}. In principle, RAI works call for proactive considerations of ``how things can go wrong''---risks, harms, and ethical issues in general---and addressing these issues proactively during development, rather than reactively after deployment~\cite{rakova2021responsible,passibarocas2019, amershiplannigtofail}. However, recent work studying how practitioners deal with RAI issues on the ground, including AI fairness~\cite{madaio2022assessing,holstein2019improving,deng2022exploring, passibarocas2019}, transparency~\cite{liao2020questioning,hong2020human,amy2022}, accountability~\cite{raji2020closing}, and overall harms mitigation~\cite{Rakova2021, amershiplannigtofail}, report that practitioners grapple with tremendous challenges ~\cite{PetersRAI}.  Proactively anticipating potential harms for complex systems deployed in heterogeneous social contexts is inherently challenging ~\cite{ShelbySociotechnicalHarms}.

To tackle this challenge, prior work called for inter-disciplinary approaches that involve technical experts, ethics experts, social scientists, and user experience (UX) professionals~\cite{metcalfimpact2021,Passimaking, Passitrust,Kashfi2017, Kayacik2019}, and multi-stakeholder approaches by also involving end-users and impacted communities to align RAI with their values~\cite{delgado2021stakeholder,sloane2020participation,halfaker2020ores}. Research also aims to provide structured yet flexible ``scaffolding'', such as guiding frameworks~\cite{gebru2021datasheets,mitchell2019model,ehsan2021expanding}, checklists~\cite{madaio2020co}, processes~\cite{liao2020questioning,raji2020closing} and tools to lower the barriers and empower individuals to engage in RAI practices~\cite{madaio2020co,liao2020questioning,amy2022}. Recent work emphasizes the need to strategically introduce friction in RAI practices to promote critical reflection~\cite{ravoka_2022}. Meanwhile, the challenges for RAI practices must also be addressed at the organizational level, by creating incentive structures and organizational frameworks, strategies and workflows that prioritize RAI practices~\cite{Rakova2021, samirthesis, vorvoreanu2023responsible}.

Aligning with the goal of providing actionable guidance for practitioners to take a proactive and multi-stakeholder approach to RAI, our proposed design process for seamful XAI tackles the challenging work of ``anticipation'' ~\cite{BreyAnticipatory} by offering two complementary perspectives to existing RAI work. First, seamful design encourages practitioners to systematically examine the \textit{sources} of potential harms in the development lifecycle rather than requiring them to fully anticipate all possible harms in the deployment context. This offers a more tractable and actionable approach to RAI. Second, seamful design prioritizes user understanding and their agency to take actions, rather than deterministic mitigation in development, acknowledging that despite the best intention, technology developers may not be able to anticipate all user needs and take actions on behalf of the users. Thus, our work also aims to introduce a new perspective and an actionable approach to this currently formidable RAI goals of anticipating and mitigating AI breakdowns.

\section{Seamful XAI Design Process}\label{sec:sxai design process}

Our goal is to adapt the concept of seams to the context of XAI systems to augment explainability and support user agency. We do so by developing a \textit{design process of Seamful XAI}, which allows us to solidify the conceptual transfer and enable stakeholders to engage in seamful XAI design. 

In the context of AI, we conceptualize seams as \textit{mismatches and cracks between assumptions made in designing and developing AI systems and the reality of their deployment contexts}. To develop the design process, we draw on key notions from the seamful design literature (reviewed in Section~\ref{seams-review}): 1) seams are inevitable, arising from the integration of heterogeneous sociotechnical components during technology deployments. Seams are revealed through system \textit{breakdowns}. 2) Instead of treating seams as problematic negatives to be erased, they can be leveraged by \textit{strategically revealing} (and concealing) seams to users. 3) The goal of this strategic revelation (and concealment) is to support \textit{user agency} (actionability, contestability, and appropriation). Our process encourages AI practitioners to take proactive and multi-stakeholder approaches to anticipate seams and design with them to mitigate potential AI breakdowns in deployment. 

Our process offers a few key benefits to complement existing RAI approaches. First, what makes it seam\textit{ful} lies around empowering users to turn seemingly negatives into positives---i.e., leverage the seams in an opportunistic manner to achieve user goals in an informed manner. Second, the process not only empowers generation but also filtering (many RAI techniques only focus on the generative part). Through the process, participants can not only anticipate seams (mismatches) from certain breakdowns but they can also figure out \textit{what to do with them}---strategically filter which seams to show and which to hide, making the design space tractable. This addresses a common weakness of anticipatory methods that are  excellent at generating risks but can create situations that overwhelm practitioners when it comes to knowing what to do with the anticipated harms. Finally, a key feature of this process is its interconnected and tightly coupled nature: not a single seam is untethered. Every seam can be traced back to a breakdown, its source in the AI lifecycle, and what user goal it serves if revealed.

\begin{figure}[t]
    \centering
    \includegraphics[width=0.8\columnwidth]{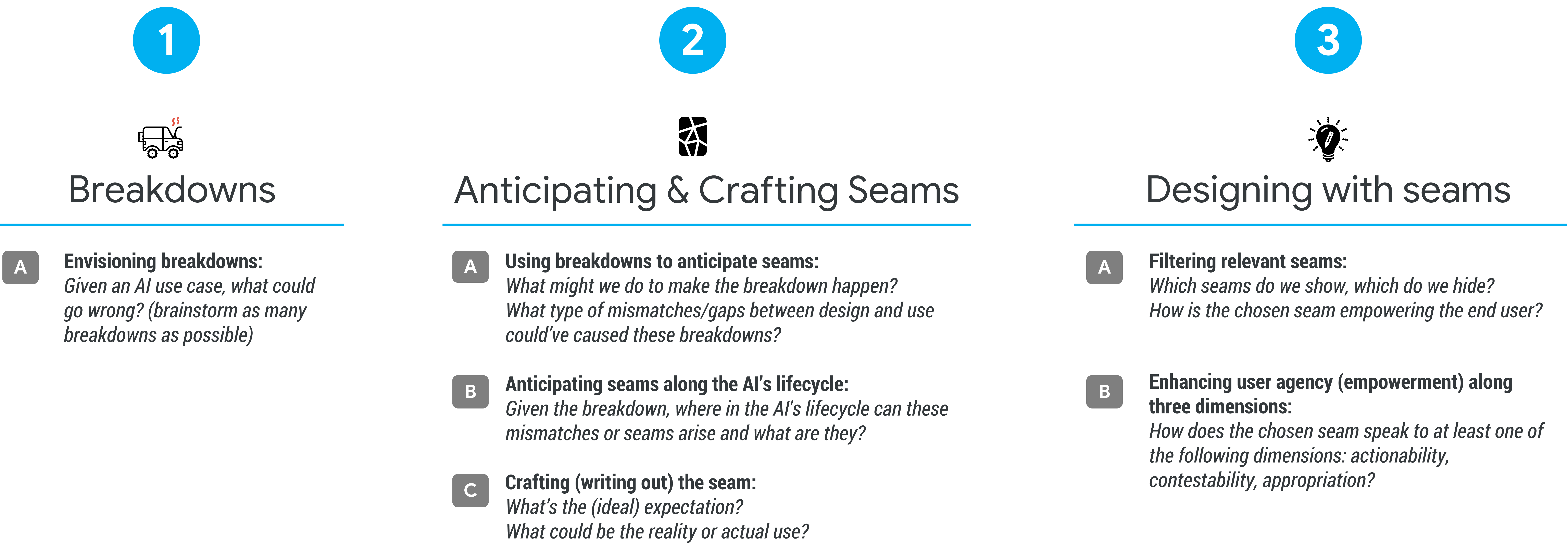}
    \vspace{-1em}
    \caption{An overview of the Seamful XAI design process with key questions relevant to each step.} \vspace{-1.5em}
    \label{fig:process123steps}
    \Description[figure]{Figure 1: An overview of the Seamful XAI design process with key questions relevant to each step. The image has three columns with different line items under each of them. First, breakdowns which has one procedural item - that of "Envisioning Breakdowns" - given an AI use case, what can go wrong? Second, anticipating and crafting seams that has three procedural items - that of using breakdowns to anticipate seams (e.g., what can we do to make the breakdown happen?), of anticipating seams along the AI's lifecycle (e.g., given a breakdown, where in the lifecycle can it emerge from?), and of crafting the seams (e.g., what is the ideal expectation and what could be reality?). Third, designing with seams which has two procedural items - that of filtering relevant seams (e.g., which seams to show or hide?) and enhancing user agency (i.e., how does the chosen seam speak to at least one of the following dimensions - actionability, contestability, and appropriation?).}
\end{figure}

Broadly, our design process has three steps (Fig.~\ref{fig:process123steps}): envisioning breakdowns, anticipating the seams, and designing with seams. We grounded the development of the process using theoretical and empirical means. Conceptually, the diverging and converging elements of the ``diamonds'' in human-centered design served as a guiding design lens. Step 1-2 could be thought of as the generative or diverging (“<” of the diamond) part whereas 2-3 is the filtering or converging (“>” half of the diamond). Another theoretical inspiration was TRIZ (Theory of Inventive Problem Solving) framework~\cite{ilevbare2013review}(elaborated in \ref{subsec: step 2 of design process}) whose adversarial thinking substantially helped participants get traction in the generative task without getting overwhelmed. Stemming from the ethos of \textit{Jugaad} (Frugal) Innovation~\cite{radjou2012jugaad}, the aim is to be resourceful and strategically use the seams as part of the solution instead of trying to hide them.
Beyond the conceptual grounding, we ran 4 focus group discussions (FGDs) with 11 AI practitioners and researchers from 4 technology companies and 2 universities. In total, we had 17 total discussion sessions within the research team punctuated by the 4 FGDs that helped us strike a balance between depth and breadth. For instance, we worked with the participants to iterate on the number of lifecycles (e.g., since seams predominantly appear in deployment, 3 slots are reserved for post-deployment stages in step 2). Once we reached a stable version of the process, we ran 4 pilot studies to further refine the process.

Below we propose how each step should be performed and share rationales behind our proposal. In Fig.~\ref{fig:mural_design_process}, we present the whiteboard used in our study to facilitate the process to provide a visual illustration for these steps. These steps should be interpreted as a suggested starting point to engage in designing seamful XAI instead of a prescriptive guidance. In this section, we use the term ``stakeholders'' to broadly refer to practitioners and end-users. Even though in the empirical study we mainly focused on practitioners’ perspective, this process can also involve end-users in a participatory manner. For example, they can be involved in the same session with the practitioners to engage in a dialogue to co-anticipate and design with seams.

\subsection{ Envisioning Breakdowns (Step 1)} 
The design process begins by asking the stakeholder ``what could go wrong?'' in their use case, drawing on their knowledge about the target users and the deployment context. The goal is to articulate multiple types of breakdown (example breakdowns are shown in Part 1 of Fig.~\ref{fig:mural_design_process}). To facilitate brainstorming, stakeholders should consider variations in user characteristics and usage contexts instead of a ``hero persona'' or the ideal ``golden path''~\cite{hong2021planning} of user workflow. 

Starting the process with breakdowns is motivated by the fundamental notion in seamful design literature that seams are revealed through breakdowns~\cite{Chalmers2003seamful, inmanribes2019, BrollBenford2005}. Articulating concrete breakdowns will offer anchoring points to trace back to what types of seams can be revealed in the next step. Starting with this step will also ensure the identified seams are intentionally grounded in potential breakdowns that the seamful XAI design should aim to mitigate. 

\subsection{Anticipating \& Crafting Seams from Breakdowns (Step 2)} \label{subsec: step 2 of design process}

This is a generative step to produce rich seams that can be filtered and designed with in the next refining step. To effectively generate seams, we propose three scaffolds: using the \textit{Breakdowns} to anticipate the seams, tracing the seams through \textit{Stages in the AI Lifecycle}, and \textit{Prompts for Crafting} (writing out) the seam for system users (Part 2 in Fig.~\ref{fig:mural_design_process}).

\subsubsection{Using breakdowns to anticipate seams}
The stakeholder will be asked to engage in an adversarial role playing---given the breakdown anticipated in Step 1, ``what might we (as developers, designers, researchers, etc.)  do to make the breakdown happen?'' In other words, what type of mismatches in assumptions between design and use can lead to these breakdowns? The answer to this question will produce a list of seams.

The adversarial thinking is motivated by the lenses of Anticipatory Failure Determination (AFD)~\cite{chybowski2018applying}, which is a part of the TRIZ (Theory of Inventive Problem Solving) framework~\cite{ilevbare2013review}. Used in high-stakes domains like nuclear engineering to proactively mitigate failures, AFD `inverts' the problem to make the failure and the search for its underlying reasons a goal directed task~\cite{thurnes2015using}. Once we have the causes for failures, we can then proactively search for solutions to prevent them. Thus, it provides a tractable structure for the challenging task of anticipating seams.
 
\subsubsection{Anticipating Seams along stages of the AI's lifecycle}
\begin{figure}[t]
    \centering
    \includegraphics[width=0.8\columnwidth]{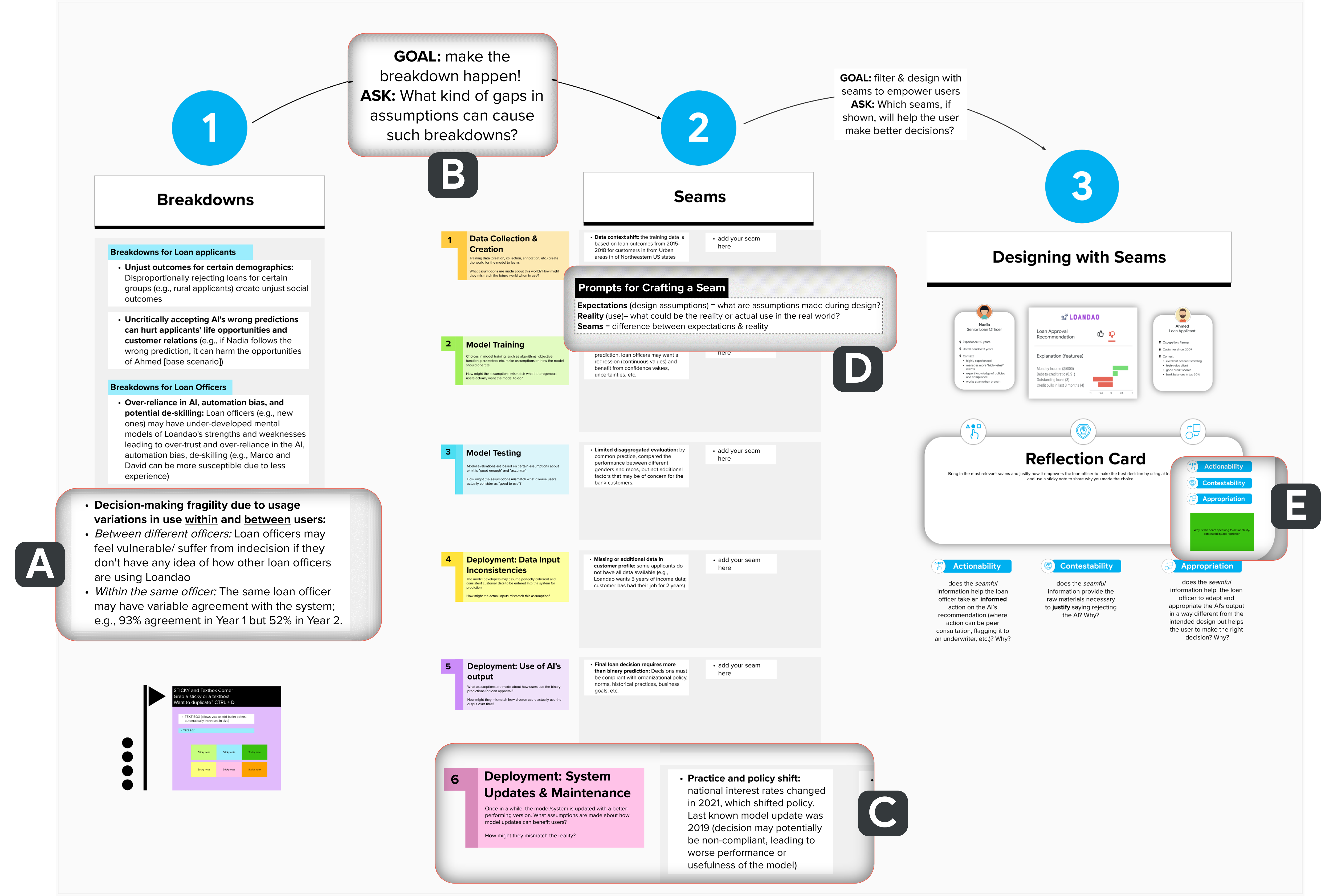}
    \vspace{-0.5em}
    \caption{A screenshot of the virtual whiteboard used for the seamful XAI design activity in the study, with zoomed-in examples. \textbf{Area 1:} Envisioning breakdown (Step 1). In the study, we provided sample breakdowns (\textbf{A}), which participants could either use directly or get inspiration for their own envisioning. \textbf{Area 2:} Anticipating \& crafting seams (Step 2). We provided guiding prompts (\textbf{B}) for effectively crafting the seams. We also shared exemplary seams (\textbf{C}) for each stage of the AI lifecycle framework. \textbf{Area 3:} Designing with seams (Step 3). We asked participants to articulate their reasoning for choosing a seam and tag which user goals the selected seam (\textbf{E}) can support for augmenting user agency. (Appendix~\ref{sec:appendix_high_res_whiteboard} has a non-annotated higher resolution version of this picture.)
    } 
    \label{fig:mural_design_process}
    \Description[figure]{Figure 2: A screenshot of the virtual whiteboard used for the Seamful XAI design activity in the study, with zoomed-in examples. There are three main areas highlighted. Area 1: Envisioning breakdowns (this is step 1). In this study, we provided sample breakdowns (highlighted in the A box - e.g., decision making fragility due to usage variations between users). Participants could either use these directly or inspire their own envisioning. Area 2: Anticipating and crafting seams (this is step 2). We provided guiding prompts (shown in B box - e.g., Goal is to make the breakdown happen and the ask is to identify which kinds of gaps in assumptions can cause such breakdowns). We also provided exemplary seams (shown in box C - e.g., shifting practices and policies).  Area 3: Designing with seams (this is step 3). We asked participants to articulate and tag which user goals the selected seam (shown in box E - e.g., actionability, contestability, or appropriation) can support for augmenting user agency. The virtual whiteboard is organized as a tripartite table with a combination of text and icons.}
     \vspace{-15pt}
\end{figure}

To further guide the anticipation of seams with actionable steps,  we ask stakeholders to think through stages of an AI lifecycle (shown in part 2 of Fig.~\ref{fig:mural_design_process}): data collection and creation, model training, model testing, data inputs in deployment, use of outputs, and updates and maintenance. ``Where in the AI's lifecycle can these mismatches or seams arise and what are they?'' In each stage, if applicable, practitioners should pinpoint assumptions that they made in the development or design that can cause the given breakdown.

Using an infrastructural perspective~\cite{leonardi_theoretical_2013,vertesi2014seamful}, our proposed AI lifecycle stages consider common technical infrastructures in model development and ask stakeholders to trace their ``seams'' with the infrastructures in deployment. To capture the sociotechnial elements,  we went beyond the traditional AI lifecycle stages (data collection, model testing, and training) and expanded into the situated use of AI (e.g., input into the AI, use of the AI’s output, and maintenance). The formation of these stages are informed by HCAI research studying AI development practices~\cite{suresh2021beyond, ehsan2021operationalizing, dhanorkar2021needs} and four in-depth discussions with industry and academic experts on seamful design and XAI. 

\vspace{-5pt}

\subsubsection{Using prompts to craft the seam}
Once stakeholders locate a seam with the above two scaffolds, they will need to craft the seam in a manner that end-users can use to support their decision-making. To craft the seam, we suggest the following prompts for scaffolding: stakeholders should ask ``What are the expectations during design?’’; next, they should consider: ``What could be the reality or actual use in the real world?’’. \textit{The seam can be written down as the difference between the expectation and reality} (the expectation vs. reality equation). Formative pilots showed that these prompts not only simplify the generation process but also make the generated seams more actionable and consistent across participants. 
This scaffold is motivated by the goal of seamful design---merely revealing seams is not enough; it has to be intentional in supporting user agency. 

\subsection{Designing with Seams (Step 3)}
\label{subsec:step3_designing with seams}
After the generative step of anticipating and crafting seams, the last step is to filter and design with them---to allow ``strategic revelation and concealment’’ by selecting the most relevant seams that can empower end users. This filtering part can be thought of as the converging aspect (the ">" of the diamond of user-centered design). This step aligns our process with the goal of seamful design: supporting user agency. We ask stakeholders to consider: to empower the user to make better decisions and mitigate fallouts from the potential breakdowns identified in Step 1, ``which seams are relevant to show? Which seams are acceptable to hide?'' 

Here we suggest another scaffold to help stakeholders to effectively filter which seams can support user agency---a \textit{Reflection Card} that ``houses'' the seams (Fig.~\ref{fig:mural_design_process}).  While it’s not the only way to present seamful designs, this card serves as a design artifact that can be shown directly to users and weave seams into the system interface.

To guide the filtering process in a deliberative manner, we unpack the concept of “user agency” into three user goals based on seamful design and XAI literature. The guiding questions below can help stakeholders decide how the seams can address agency through supporting these goals. 
\begin{itemize}[nolistsep,noitemsep,left=0.5em]
    \item Actionability: Does the seamful information help the user take informed actions on the AI’s recommendation? Why?\looseness=-1
    \item Contestability: Does the seamful information provide the resources necessary to justify saying no to the AI? Why?
    \item Appropriation: Does the seamful information help the user to adapt and appropriate the AI's output in a way different from the provided design but help the user? Why?

\end{itemize}

\noindent
Stakeholders are asked to articulate how a selected seam can address at least one of the three aforementioned user goals. If the seam does not address the goals, it can be concealed. To promote traceability and facilitate downstream design decisions, they will tag the selected seams with the targeted user goals (stickers in the Reflection Card in Fig.~\ref{fig:mural_design_process}). 

% \vspace{-8pt}

\noindent
\textbf{Practical suggestions for the design process:}
\textit{When generating the seams}: Instead of doing it collectively, the process can be performed individually by different stakeholders. We recommend to have one facilitator serving as the junction point for these conversations. This alleviates the challenges of getting multiple teams synchronously, mitigate conformity pressure, and group thinking. Stakeholders can also engage in brainwriting~\cite{vangundy1984brain} where seams are generated separately followed by a discussion to refine things. 

\textit{When filtering seams}:  When many seams are considered relevant for supporting user agency, some prioritization strategies can be utilized, such as ranking the the seams based on how many dimensions of user goals they address (e.g., by the number of stickers they had). Stakeholder can also engage in an additional prioritization step, such as by evaluating the alignment of a seam with business goals or organizational values.  

\textit{Post-deployment}: The process is designed to be a ``living'' one,  used both pre- and post-deployment. Given not all downstream harms might be proactively anticipated (e.g., in large-scale AI systems), teams can take multiple passes (e.g., quarterly) throughout the product's lifetime.

\section{Methods: Empirical Study to Apply the Design Process}
\label{mehods}
Using our proposed design process described above, we conduct an empirical study with AI practitioners. The study is intended to be both a formative validation of the proposed method and to gain empirical insights about how seamful XAI design can help AI product development practices, especially in terms of augmenting user agency and explainability. This study also enables us to reflect on---and improve---the proposed process.

To learn from a wide range of perspectives, we chose to conduct an interview study that centers around a scenario-based design activity. A scenario-based design (SBD) allows simultaneously suspending the need to specify high-fidelity technical details while ``envisioning future use possibilities” through immersive storytelling~\cite{rosson2009scenario}. Participants went through the design process outlined above (in Sec.~\ref{sec:sxai design process}) with a given AI product scenario.  Following the design activity, we interviewed them as they critically reflected on the process including: (a) the transferability of the design process to product contexts that they were familiar with, (b) how seamful design can impact AI explainability and user agency in their contexts, and (c) the effectiveness of the process. Below we begin by introducing the scenario we used and how we developed it, then describe the procedure, recruitment and qualitative data analysis.

\subsection{Developing the Scenario for the Design Process }
We need a scenario of an AI use case, which should include both user personas\footnote{In our study, when we refer to the `user', we mean the loan officer (not the applicant) who is the user of Loandao. The seamful information thus should be crafted such that it helps the officer make their decision.} and usage contexts. It should balance two critical aspects: first, it should be concrete enough to generate design ideas, and second, it should afford enough malleability for participants to co-opt it.

We chose \textit{lending} as the target domain for the scenario because of its broader relatability and consequential nature. This choice was informed by 7 consultation sessions with 18 diverse stakeholders (AI and HCI researchers, practitioners, and users) where we explored multiple possible scenarios including cybersecurity, radiation oncology, lending, and hiring. We found that even for people who did not work in the lending domain, everyone nonetheless had some level of lived experience with loans. Participant feedback also suggested that seams revealed in lending situations had the potential to be impactful for both applicants and loan officers.

\begin{figure}[t]
    \centering
    \includegraphics[width=0.8\columnwidth]{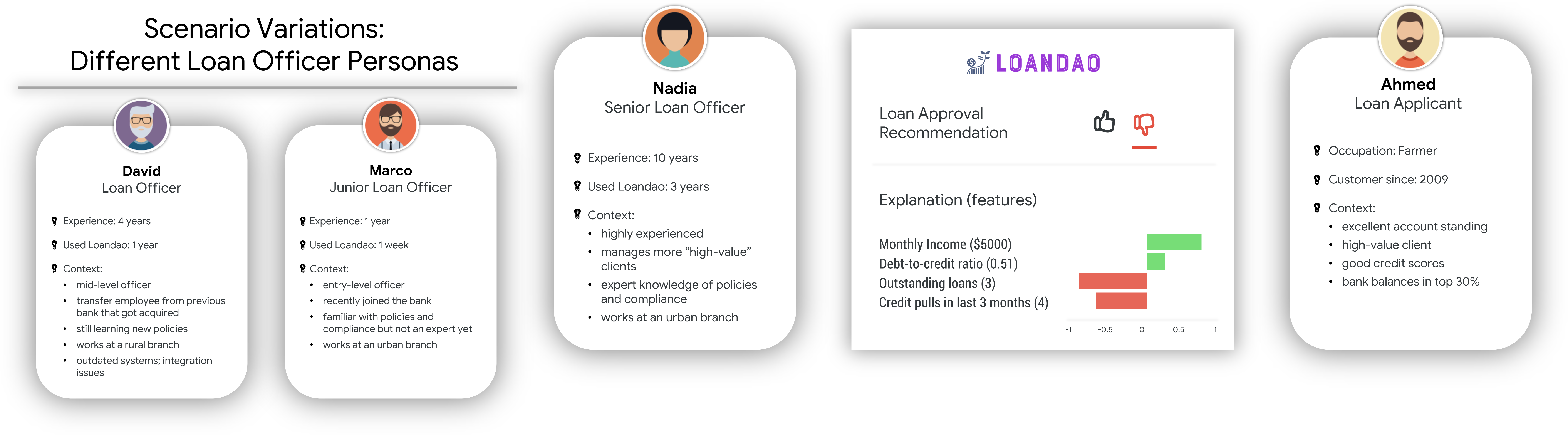}
    \vspace{-0.5em}
    \caption{A visual presentation of the lending scenario used in our study showcasing the ``backstory'' of each persona. The different loan officer personas were used by participants to think of variations in the scenario to generate new breakdowns and seams.} 
    \label{fig:scenario}
    \Description[figure]{Figure 3: A visual representation of the lending scenario used in our study showcasing the "backstory" of each persona. The image has two parts. On the right side, we have the main scenario setup consisting of three columns. The first column shows the persona of a senior loan officer (Nadia) with little bits of information. The third column shows the persona of the loan applicant (Ahmed) with little bits of information. The second second column shows the AI's decision as it would appear in the Loandao system. There is a loan approval recommendation - a thumbs up or down icon. This is followed by a short explanations of the features that the AI model found useful in making the prediction. On the left side of the image, there are two other loan officer personas that participants could use to think of variations in the scenario to generate new breakdowns and seams. These personas include a junior loan officer (Marco) and another loan officer (David). Both these personas also contain little bits of accompanying information.}
    \vspace{-10pt}
\end{figure}

We developed our scenario by basing it on real-world use cases as informed by domain experts’ insights and feedback. We did so by having iterative conversations with six loan officers who had prior experiences with automated lending support. All participation was voluntary and through personal rapport. Specifically, we constructed the narrative in our lending scenario through nine conversation sessions, where each session involved at least three loan officers. We asked the loan officers to share stories about real-world mismatches between idealized AI design and its use, which helped us generate example seams to show to the SBD participants; a list of realistic breakdowns is shown in Fig.~\ref{fig:mural_design_process}. We co-constructed not only the narrative but also developed the different personas involved in the story with the loan officers. The loan officers also shared a realistic list of feature-level explanations provided by the systems they currently use, which we used to ground the AI system introduced in the SBD scenario as well. Finally, the loan officers validated the usefulness of the exemplar seams we generated to seed the brainstorming with our study participants later. Below we share the backstory where all names are pseudonyms (captured visually in Fig.~\ref{fig:scenario}):
\begin{quote}
    Nadia, a senior loan officer, is using Lonadao, a lending decision-support software, to evaluate if Ahmed, the applicant, should get a loan or not. Despite Ahmed being one of the bank's best customers, Loandao rejects his loan application. When Nadia checks Loandao's feature-level explanations, she cannot find anything that leads her to question Loandao's rejection. She is in a tough spot: she knows Ahmed has a good track record with the bank, yet the AI has rejected him. Should she accept the AI's decision? What else does Nadia need to know to make an informed decision?
\end{quote}{}
 
\noindent 
To encourage stakeholders to consider variations in user characterstics and usage contexts, as suggested in Step 1 of our process, we add two additional loan officer personas, Marco and David, who vary in both organizational and technological experience from Nadia (Fig.~\ref{fig:scenario}). For instance, some participants in the study considered Marco's use of Lonadao and found new seams by asking: what type of seam would be more relevant to a less experienced officer?

\subsection{Study Procedure}
We conducted the study during the Covid-19 pandemic using video conferencing and online whiteboarding tools.
Lasting 65 minutes on average, our study activity consisted of \textit{three} phases.

The \textit{first} phase is the introduction. After providing informed consent, participants watched an orientation video that covered the lending scenario and the personas, provided a high-level idea of the concept of seams as mismatches between design and use, and why seams can help augment agency and explainability. We then answered clarifying questions from participants. Next, we showed participants the Reflection Cards (where seams will be presented to users) and primed participants to think of the card as expensive real-estate: a seam had to \textit{earn its place} to be there. This priming helped participants craft seams intentionally later on. We also primed them to think of examples in their own domain where such a process could apply (a point we returned to in the third phase of the study). Then, we provided an end-to-end example walkthrough of the steps of design activity (Fig.~\ref{fig:mural_design_process}): selecting one of the breakdowns, identifying and crafting the seams for it, and then transferring the seams to the Reflection Card and arguing for why that seam addressed at least one of three dimensions of user agency: actionability, contestability, or appropriation.

The \textit{second} phase is the design activity, following the design process outlined in Sec.~\ref{sec:sxai design process} and applying it to the lending scenario. Here, the participant led the activity and we (as facilitators) guided them. To seed brainstorming, we provided four examples of realistic breakdowns in AI-powered lending and exemplar seams, all informed by discussions with the loan officers as described in the last section. Participants started with breakdowns (Step 1), followed by anticipating and crafting seams (Step 2), and then going through a filtering process to design with the seams (Step~3). 

The \textit{third} phase is an interview reflecting on the design activity and transferring it to their own product contexts. We started with transfer cases: we asked if participants could share an example in their domain where they could apply seamful design, for instance as a  reframing of a prior project, or a speculative example in their domain. Once participants found an example, we asked them to highlight the breakdowns, the seams, and how awareness of the seams would impact the user experiences of their products. Next, participants shared their overall thoughts on the effectiveness of the design process: what worked well and why. Finally, participants shared their final thoughts on how thinking seamfully affects the explainability of the system, how seamful information can empower users, and what challenges they faced as well as suggestions for improvements.

\subsection{Study Recruitment}

\begin{table}[t]
\sffamily
\caption{Details on study participants: Organizational roles and experience levels.}
\label{table:participant details}
  \begin{minipage}[t][][t]{0.55\columnwidth}
  \centering
    \footnotesize
	\setlength{\tabcolsep}{5pt}
    \begin{tabular}{lp{4cm}r}
 \textbf{ID}&\textbf{Organizational Role}&\multicolumn{1}{c}{\textbf{Experience}} \\
 \toprule
 
  \rowcollight \multicolumn{3}{c}{\textit{AI Research}}\\
P28   & Academic Researcher (XAI)  & \textgreater 3 ys.\\
P29   & Academic Researcher (XAI)  & \textgreater 4 ys.\\
P31  & Academic Researcher (XAI) & \textgreater 10 ys.\\
P32   & Academic Researcher            & \textgreater 7 ys.\\
P33   & Academic Researcher (XAI)  & \textgreater 4 ys.\\
P07   & Industry Researcher (XAI)  & \textgreater 5 ys.\\
P21   & Industry Researcher              & \textgreater 4 ys.\\
P24   & Industry Researcher              & \textgreater 10 ys.\\
P34   & Industry Researcher (XAI)  & \textgreater 4 ys.\\

\rowcollight\multicolumn{3}{c}{\textit{AI Policy \& Governance}}\\
P10  & AI Ethics/ Responsible AI Expert & \textgreater 3 ys.\\
P11  & AI Ethics/ Responsible AI Expert & \textgreater 5 ys.\\
P19  & AI Ethics/ Responsible AI Expert & \textgreater 10 ys.\\
P20  & AI Ethics/ Responsible AI Expert & \textgreater 10 ys.\\

 \rowcollight \multicolumn{3}{c}{\textit{Engineering and Data Science}}\\
P13   & Data Scientist              & \textgreater 5 ys.\\
P16   & Data Scientist              & \textgreater 10 ys.\\
P17  & Data Scientist               & \textgreater 4 ys.\\
P18  & Data Scientist               & \textgreater 5 ys.\\
P40   & Data Scientist (Lending focus)             & \textgreater 3 ys.\\
P41   & Data Scientist (Lending focus)             & \textgreater 3 ys.\\
P42   & Data Scientist (Lending focus)             & \textgreater 4 ys.\\
P43   & Data Scientist (Lending focus)             & \textgreater 7 ys.\\
% & & \\

% & & \\
\bottomrule
\end{tabular}
  \end{minipage}\hfill
    \begin{minipage}[t][][t]{0.44\columnwidth}
  \centering
    \footnotesize
	\setlength{\tabcolsep}{4pt}
    \begin{tabular}{lp{3.5cm}r}
 \textbf{ID}&\textbf{Organizational Role}&\multicolumn{1}{c}{\textbf{Experience}} \\
 \toprule
 \rowcollight\multicolumn{3}{c}{\textit{Product Development \& User Research}}\\
P02   & Product Manager  & \textgreater 5 ys.\\
P03   & Product Manager   & \textgreater 4 ys.\\
P11   & Product Manager  & \textgreater 10 ys.\\
P14   & Product Manager  & \textgreater 5 ys.\\
P19   & Product Manager   & \textgreater 8 ys.\\
P22   & Product Manager  & \textgreater 15 ys.\\
\hdashline
 % \rowcollight\multicolumn{3}{c}{\textit{User Research}}\\
P08   & AI-UX Designer  & \textgreater 15 ys.\\
P12   & AI-UX Designer  & \textgreater 10 ys.\\
P15   & AI-UX Designer  & \textgreater 3 ys.\\
P26   & AI-UX Designer  & \textgreater 5 ys.\\
P27   & AI-UX Designer  & \textgreater 4 ys.\\

\hdashline

P01   & UX Researcher  & \textgreater 5 ys.\\
P04   & UX Researcher  & \textgreater 5 ys.\\
P05   & UX Researcher  & \textgreater 5 ys.\\
P06   & UX Researcher  & \textgreater 5 ys.\\
P20   & UX Researcher  & \textgreater 5 ys.\\
P23   & UX Researcher  & \textgreater 5 ys.\\

 \rowcollight\multicolumn{3}{c}{\textit{(End) Users}}\\
P35   & Loan Officer  & \textgreater 6 ys.\\
P36   & Loan Officer  & \textgreater 8 ys.\\
P37   & Loan Officer  & \textgreater 3 ys.\\
P38   & Loan Officer  & \textgreater 10 ys.\\
P39   & Loan Officer  & \textgreater 7 ys.\\
\bottomrule
\end{tabular}
  \end{minipage}
\Description[Table showing participant details]{Table 1: Details on study participants consisting of their organizational roles and amount of experience. This is organizes as a table with three columns: participant ID, their organizational role, and years of experience. There are five categories of participants based on their roles: AI research, AI policy and governance, engineering and data science, product development and user research, and end users.}
\end{table}

Our aim was to gather multi-disciplinary perspectives from AI stakeholders, including researchers, practitioners, and end-users of AI-powered applications. We recruited through online ads with a sign-up form survey (advertised via email and social media), which gathered 258 total submissions. In this survey, we asked for the following: experience with AI systems, experience with XAI, their role, domain,  geographic location, and number of years of experience, along with questions where they can provide examples. For screening, we balanced the following factors: range of work experience (from early career to experienced), XAI experience, variety of domains, geographic diversity, project timelines and budget, and most importantly, scheduling constraints across multiple time zones. We prioritized participants who provided detailed work examples, which showcased their motivation to engage.Our final pool comprised of a diverse group of 43 participants comprising of Product Managers, Data Scientists, UX Researchers, Designers, AI Ethics/Responsible AI experts, Researchers (both Academic and Industry-based), and Loan Officers. Table~\ref{table:participant details} provides a breakdown of their roles, domains, and experience. Each participant was provided US \$50 as an appreciation for their time. 24 out of 43 participants self-identified as female while the rest identified as male. While the majority of participants were located in the US (22 out of 43), we had participants from Australia, Belgium, Canada, Denmark, Germany, India, Netherlands, and the United Kingdom.  Not only do we have Loan Officers representing their voices as potential end users of the system, we also recruited a few Data Scientists who specialized in developing automated lending systems.

\subsection{Qualitative Analysis}
 
Overall, we analyzed 2796 minutes of transcribed videos from the 43 study sessions as informed by the grounded theory approach~\cite{Charmaz2014, Strauss1990}. Taking an inductive approach, we started the process with an open-coding scheme and iteratively produced in-vivo codes (generating codes directly from the data). One researcher performed the iterative coding, punctuated by frequent discussions with the research team where we constantly compared and contrasted the codes, refining and reducing the variations in each round. We referred to the videos and whiteboards as needed. Next, we analyzed the data using axial codes, which involves finding relationships between the open codes and clustering them into different categories (e.g., `revealing the AI's blind spots').  Finally, we unified the axial codes and consolidated them to selective codes (e.g., `enhancing explainability'). We discuss the selective codes as the main themes in our findings section where the axial codes are bolded (ones that add to the theme), wherever applicable.

\begin{figure}[t]
    \centering
    \includegraphics[width=\columnwidth]{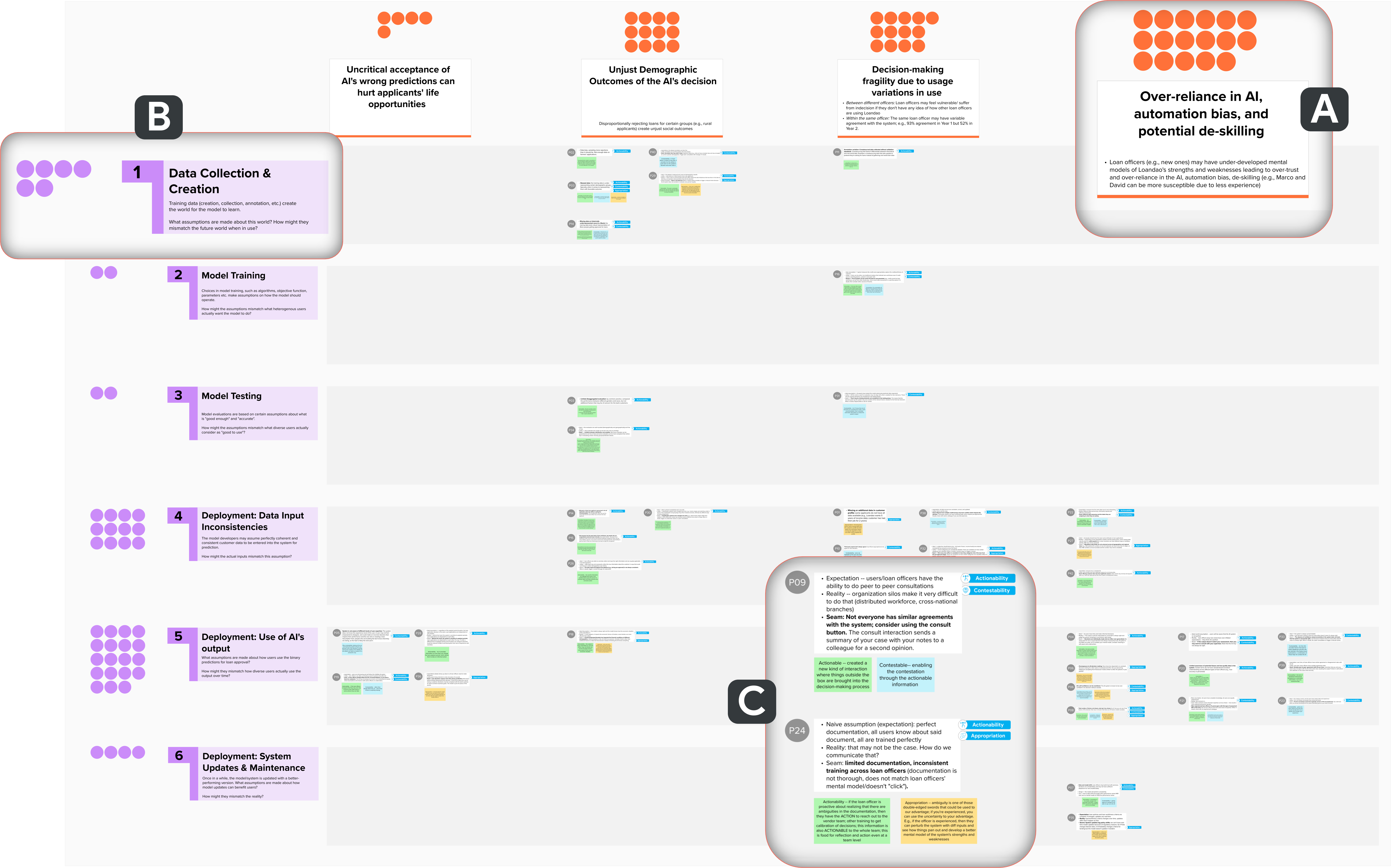}
    \vspace{-0.5em}
    \caption{A bird's eye view of all seams from all participants for all breakdowns along all AI lifecycle. \textbf{A:} the dots above this breakdown provide the number of times it was chosen (e.g. 17 times). All the seams appearing below this column were crafted in connection to this particular breakdown. \textbf{B:} the dots showcase the number of seams crafted for a given AI lifecycle stage (6 seams were crafted for this stage). \textbf{C:} a zoomed-in view to provide a sample of seams crafted by our participants along with the justifications for how the seams enhances agency. (Appendix~\ref{sec:appendix_high_res_allseams} has a non-annotated higher resolution version of this picture.)
    } 
    \label{fig:all_seams}
    \Description[figure]{Figure 4: A bird's eye view of all seams from all participants for all breakdowns along all AI lifecycle. This is in the form of a table. There are six rows signifying the different stages in the AI lifecycle: data collection and curation, model training, model testing, data input and inconsistencies, use of AI's output, and system updates and maintenance. There are dots next to the row headers that showcase the number of seams crafted for a given AI lifecycle stage (e.g., 6 seams were crafted for the data collection and curation stage). There are four columns indicating the different breakdowns identified by participants (e.g., box A highlights one such breakdown - that of overreliance on AI that was chosen for all the seams appearing below in this column. There are dots above the column header that provide the number of times this was chosen (e.g., overreliance on AI was chosen 17 times). Box C shows a zoomed-in view to provide a sample of seams crafted by our participants along with the justification for how the seams enhance agency.}
    \vspace{-10pt}
\end{figure}

We also collated all seams from all participants and clustered them along their respective breakdowns and AI lifecycle stages, creating a single snapshot of the entire study to get a sense of the distribution of breakdowns chosen and seams generated (visually depicted in Fig.~\ref{fig:all_seams}). 

\section{Findings}
In this section, we report on our key empirical findings, \textbf{organized in four parts}. \textit{First}, we describe how our design process helped participants to not only identify, craft, and design with seams in the context of the study scenario, but also transfer learning to their own product contexts. \textit{Second,} we describe the value of introducing and operationalizing the notion of seamful XAI in terms of augmenting explainability and agency goals. \textit{Third,} we describe the efficacy of our design process more broadly regarding how it helps participants proactively design for failures, make concrete abstract RAI goals, and address often amorphously-defined AI harms. \textit{Finally,} we touch upon challenges and areas of opportunities concerning our process as described by our participants.

\subsection{Design Activity Effectiveness
: Anticipating, Crafting, and Designing with Seams}
We first provide an overview of our study activities, including information about the seams generated. Then we share the context for how participants identified and designed with seams.

\subsubsection{Overview}
Our 43 participants generated a total of 85 seams. 
By following our process, \textbf{every participant was able to  identify and craft at least one seam.} This is encouraging because only one participant had some knowledge about seamful design before the study; everyone else started with no knowledge. 
As depicted in Fig.~\ref{fig:all_seams}, overall, participants filtered and chose to reveal 47 relevant seams.
For example, around dataset and model training, seams like ``skewed data'' (P03), ``data and model drift'' (P07), and ``the AI system can overly focus on one parameter such as credit score and ignore other parameters''(P16) were identified. On deployment side, participants crafted seams that looked at gaps in usage from data input as well as updating and maintenance; e.g., seams like ``reports from multiple credit bureau's may have conflicts which impacts AI's decision'' (P38), ``model updates does not guarantee compliance with policies'' (P37), and ``variable peer-to-peer agreement with AI over time'' (P39) were crafted. 
Participants placed these seams on the Reflection Card and justified why the seamful information augments user agency.

Each breakdown had at least 5 participants working on it. All AI lifecycle stages had seams crafted along them. The most popular breakdown was “Over-reliance on AI, automation bias, and potential de-skilling” (N=17) and the most popular AI lifecycle stage was “Deployment: Use of AI's prediction” (N=23). While participants added seams to all AI lifecycle stages, the majority (35 out of 47) were anticipated in deployment stages (4-6). 
No obvious differences in the number and type of seams addressed were observed for participants of different job roles.
Below we share how our process helped participants anticipate, craft, and design with seams. We then underscore how participants found the process effective in transferring the concepts to their own domains.

\subsubsection{Anticipating and crafting seams}
Participants highlighted three main reasons why the process was effective in empowering them to engage in seamful design, all of which are tied to the scaffolds our process provided. First, \textbf{the adversarial thinking during breakdown was helpful} because “the first impulse of when we think of breakdowns is to stop them from happening” (P19). “Making breakdown a goal-centered task is super effective” because as “technologist[s] [we are] often too solution oriented and don’t pause enough to think through what the exact problems are” (P07, P24). The “provocative” nature of asking `what could we do to make the breakdown happen' helped participants to “not go to the solution first” but instead “have the \textit{freedom} to really think about bad outcomes without being burdened by the hero expectations” (P03, P01, P24). 

Second, \textbf{having the AI lifecycle stages listed out helped} participants “find a home for the seams because the gaps could come from anywhere” (P25). The different stages also made the “thinking process deliberative and structured” and “showcased the sociotechnical reality of AI systems” (P29, P23). Third, \textbf{prompts for crafting were “instrumental”} to frame and write out the seam “in a productive way” (P15, P20, P43). “Coming up with the seam is one thing but figuring out how to write it in a way that is helpful to someone else is a different ball game”  (P06). 
Writing down “the [ideal design] expectations and the reality [real use]” made the process “reproducible and doable even for first-timers” (P17, P32). Having the process broken down “into bite sized pieces” made the “learning curve less steep, promot[ing] self-reliance” (P06, P07):

\begin{quote}
    No real-world technology follows the golden path. Having this expectations vs. reality equation has boiled the art of crafting a seam into a recipe that anyone can easily follow...that's powerful. (P08, a senior AI UX designer)
\end{quote}

\noindent

\subsubsection{Designing with Seams}
\textbf{Every participant succeeded to design  with the seams they crafted}: to strategically select seams that would be worth revealing to augment user agency.

This step was supported by the use of reflection cards and the breakdown of agency into three dimensions: actionability, contestability, and appropriation. Participants \textbf{appreciated how the deliberative exercise helped them design seamfully}. \textit{First,} the “decomposition of agency into the three dimensions helped them “grapple with the abstract concept of agency” (P07, P10). They underscored how the three dimensions were “more accessible, operational, and on the ground” (P15). \textit{Second,} the deliberation empowered participants to be “decisive about filtering the seams and ensuring that every seam in the reflection card had earned its place” (P33). This way, there was no “seam bloat” problem where “practitioners would jam too many seams and overwhelm users” (P39, P28). \textit{Third,} the deliberative process helped participants “embed their design voices because the seams revealed are just as intentional as the ones hidden” (P18). 

\begin{quote}
    Most RAI frameworks \textit{stop} at the \textit{generation} phase, leaving us with a list of harms and \textit{no idea of what to do with them}. This is the \textit{first one} that lets me filter and figure out what to \textit{do} with them \textit{systematically}. It's like the double diamonds of user-centered design...[steps] 1-2 being the generative part and 2-3 as the filtering part. (P06)
\end{quote}

\subsubsection{Transferring and teaching about seams}
Participants reported that the process was \textbf{teachable} and \textbf{helped them transfer the concepts to their own use cases.} As such, \textbf{every participant succeeded in the transfer task} with varying levels of depth by reframing their past projects through a seamful lens. These comments demonstrate the effectiveness of our process as all participants, except one, was introduced to the concept of seams for the first time with our 45 minutes co-design activity. 

Overall, participants found the process to be a “teachable exercise, [one that's]...generalizable to a lot of AI use cases” (P14). In particular, they appreciated “how the process converted abstract concepts like seams into specific mismatches” (P27). They also enjoyed the “simplicity of the 1-2-3 process of going from breakdowns to seams to designing with seams on the [reflection] cards” (P05). The following participant sheds light on what worked for successful transfer:

\begin{quote}
    I'm surprised how easily I could map this new lens onto my old projects...What worked really well for me was how the process was broken down to bite-sized pieces and transitions were clear...loved the adversarial thinking we did for the breakdowns. (P40)
\end{quote}

\noindent
The transfer use cases covered a diverse array of AI-powered applications in domains such as finance, healthcare (radiology, oncology, mental health), agriculture tech, self-driving cars, humanitarian efforts, and air travel.
One transfer example was around an airline algorithmic pricing gone awry during Covid-19. ``If travel agents knew [the seam] of how the model was built on pre-Covid data and it was not capturing real-time market dynamics, then they wouldn't either trust it blindly or reject it completely'' (P20). This awareness could have helped calibrate their reliance appropriately, allowing them to contest and appropriate. Another example focused on AI-powered treatment planning in Radiation Oncology (RadOnc). If RadOncs in South Asia knew the seamful information of the model being trained on North American data with large variance annotations, it would have helped them contest the AI recommendations and fight organizational pressure to ``accept the made in USA AI model” (P07). It would have helped them also appropriate the recommendations because treatment standards vary regionally; for example, in South Asia, ``hypofractinations [higher doses per fraction] are the norm due to machine and electricity black outs, things you don't have to worry in the US” (P24). 
Below is an in-depth walk-through of a transfer case about an AI-powered international aid allocation algorithm gone wrong:

\begin{quote}
Our [ideal] assumption was that if we redacted the protected categories such as race, the algorithm would be blind to these and thus be fair. However, there was model leakage [breakdown in reality] where the algorithm latched on to ZIP codes, made weird correlations, and was discriminating based on that...if we're aware of the seam of model leakage, then it'd have explained why the bias existed.
% and help address the AI's flaws. 
(P03)
\end{quote}

\noindent
This example illustrates the depth of transfer. The fact that all participants “were able to do the transfer exercise in a limited time is a testament to ease of adoption of the process” (P25).

\subsection{The Value Propositions of Seamful XAI}
In this section, we share empirical insights on the value propositions for Seamful XAI. We focus on the two goals that our work set out to approach with seamful XAI: enhancing explainability and user agency. Beyond this, we also share how our process can benefit AI product practices.

\subsubsection{Enhancing Explainability}\label{sec:enhancing explainability}
A core motivation for this work was to enhance explainability through seamful information. Participants’ responses validated this position and further demonstrated how seams can enhance explainability.
Understanding is a key part of explainability. One of the core value propositions of seams is that it exposes the user to actionable information that augments one’s understanding. There is a “natural fit between explainability and seamful design because the places where the seams occur have the highest chance of creating negative downstream impact...[which] when revealed can explain the AI’s actions in a situated manner” (P30). Below we share \textit{three} ways in which seamful information helps with explainability.

\textit{First,} seamful information provides \textbf{peripheral vision by situating AI systems in the broader sociotechnical context}. With seamful information, “our blinders are taken away and we can see more, [which] helps us to understand the AI’s working, especially its mistakes” (P43). Seams make it “clear that an AI doesn't live in a vacuum” (P04). A “model-centered, seamless view can only explain what’s going on inside the [black-] box” (P21). Revealing seams along the ``entire AI lifecycle provides “peripheral vision that can highlight AI blind spots [that] otherwise [will] be missed if we take a seamless view” (P09), further elaborated by this participant:

\begin{quote}
    For us data scientists, the AI isn’t the black box because we build it, but everything around it is...like how the data collection conditions impact my model’s performance...Having that awareness is extremely important for me to explain the AI. (P16)
\end{quote}

\noindent
\textit{Second,} seamful XAI can \textbf{reveal the AI’s blind spots, highlight its fallibility, and showcase the strengths and weaknesses of the system,} which can facilitate \textbf{calibration of trust and reliance on the AI.} Participants felt that awareness of the blind spots can “improve explainability because if I know what the system cannot do, I can better explain why it’s unable to make certain decisions” (P25). Seams also convey the information about the blind spots in “a lay friendly manner, making things more accessible to non-AI experts” (P02). 
A clearer picture of the AI’s strengths and weaknesses is essential “to know \textit{when} to trust the AI vs. not” (P07), which is one of the “most challenging things with AI because performance is never uniformly good or bad” (P23). Trust can be calibrated “from knowing the AI’s blind spots, which the seams can reveal” (P33). Having that understanding is empowering because “it can help people make better sense of AI’s output when things go wrong” (P34). This participant further encapsulates this point: ``The improvement in explainability seems to be an implicit outcome of seamful design. By showing the seam, you bust the mythology that AI is all knowing...it reduces mindless AI acceptance'' (P34).
\textit{Third,} seamful information affords a unique type of understanding and explainability by \textbf{helping users understand why the AI may not or cannot do certain things}:
% This participant captures the point effectively:

\begin{quote}
The explainability from seams is \textit{very different}...If I know there's annotator bias, I can do forward causal inference. 
It explains why the AI \textit{cannot} make fair decisions even \textit{if} everything looks good at a feature level... The usual XAI techniques tells me the ‘why-part’, but seams can help me understand the ‘why-not’ part. (P28, emphasis added)
\end{quote}

\noindent
Having information of both why and why-not can promote “holistic understanding and counterfactual reasoning mechanisms” (P24), which can improve explainability~\cite{ehsan2021expanding, keane_good_2020}.

\subsubsection{Augmenting Agency}
Participants’ comments also validated the end-goal of seamful XAI in supporting user agency, and delineated its paths to do so.

Broadly, by providing peripheral vision of the AI’s blind spots, \textbf{seamful information expands the action space of what users can do,} augmenting agency. At its core, “seams are about opening up new kinds of interactions with the system; knowing the limitations opens spaces for how humans interact with systems” (P32). Examples of user actions include new interactions in the form of “peer-to-peer consultations...contesting the AI...or appropriating the AI’s output in a way different from its intended use” (P21). Participants reflected on how the ``western obsession with perfection often leads to an unachievable seamless ideal'' (P13) and how seamfulness can be empowering: ``seamless design is a take it or leave it paradigm...user has no voice. After identifying seams, users evaluate what to do with it—having that choice is empowering.'' (P18)

Information in seams can \textbf{convert ``unknown unknowns'' to ``known unknowns''} that can empower users to know “where to start an investigation; e.g., if [they] know that the model doesn't capture a regulatory change, while [they] can't fix the model, [they] have two things— understand the failure and know where to begin [the] investigation” (P03). 
“Without seams, people may not even know what to ask for” (P07). Often AI users “automatically assume that they are at fault and AI is right”, knowing the “limitations through the seams” can convert “unknown unknowns to known unknowns, which can empower [users] to contest the AI.'' (P05, P07, P34)

In our study, \textit{we scaffolded and operationalized user agency along three dimensions of user goals with a decision-support AI system: \textbf{actionability, contestability,} and \textbf{appropriation}}. When participants designed with seams (step 3 in Fig.~\ref{fig:mural_design_process}), they also articulated how the chosen seams augment user agency. Below we synthesize participants’ justification along with concrete examples of user actions.

\textit{Actionability} comes to the forefront when we focus on what people can \textit{do} with the information in the seams. A major way in which seams addressed actionability was when \textit{the seamful information empowered the user to “act on it in an informed manner”} (P08). 
For instance, the action could be “accepting the AI’s decision, clearly knowing why it’s appropriate,” “flagging [it] for a peer consultation to double check [the user’s] intuition,” or “recommend[ing] customers to correct their applications or credit reports if mismatches around data inconsistencies show up” (P19, P40, P39). 

\textit{Contestability} becomes relevant when the AI’s decisions “go against users’ intuition” or are “surprising” (P36, P21). Contesting a decision is a critical path of exercising user voice in algorithmic decision-making~\cite{aizenberg2020designing,lyons2021conceptualising,selbst_fairness_2019}. One of the key themes around contestability was when the seamful information provided \textit{justifiable information to say no to the AI, allowing the user to exercise their domain expertise} instead of blindly trusting the AI. 
For instance, if stakeholders “are aware of data drifts that produce biased decisions” or that “prior users rejected the AI”, it gives the “right ammunition to contest the AI despite organizational pressures to use it” (P12, P30, P27).

\textit{Appropriation} supports the ``how'': when users are dissatisfied with the current system, “\textit{how} to reinterpret or override the system” (P10, emphasis added). This dimension has an explicit “shift in the focus towards the human as the main driver” (P22). 
Seams facilitated appropriation when \textit{they afforded information that empowers the user to “co-opt”, “re-interpret”, or “re-work” the system to “aligns with the user’s goal”} (P03, P29, P41, P28). 
For instance, if participants knew of the “mismatch where masking of protected categories (such as race) negatively impacted the AI’s decision, [they] could fiddle with the input to produce fairer decisions” (P36). 
Appropriation can be thought of as “feedback to the system, a way of sharing their voice back” (P40).

Notably, every justification touches on a combination of technology, infrastructure, and work practices that encompasses the AI lifecycle. Thus, participants were able to “view” the AI in its situated environment, addressing agency by explicitly considering the human-AI assemblage.

\subsubsection{The dual value proposition: revealing seams while proactively mitigating harms}\label{sec:dual value prop}
Our design process can be a \textbf{resourceful way to not just reveal seams but also anticipate and mitigate harms.} Generating seams has an intrinsic value. The generated seams “practically become a ready-made task list of potentially harmful things to work on” (P09). Working through the list of seams can “serve as a map to help companies conduct risk assessments in a resourceful way” (P21):

\begin{quote}
There's a dual value proposition: once you’ve generated the seams, you can neutralize many of them because of proactive identification. 
Even if some seams are unfixable (e.g., regulation changes making the AI non-compliant), you can leverage them resourcefully to help with better decision-making. Thus, you get two for the price of one. (P33)
\end{quote}

\noindent
Typical “Responsible AI frameworks call for envisioning harms, which is a great start” (P03); what is \textit{unique about Seamful XAI} is that “seams aren't just shown for the sake of showing them...they are used in opportunistic ways to help the user” (P14). Thus, it empowers users to “discover the problem \textit{and} do something about it” (P34), going deeper than practices that stop at anticipating harms.

\subsection{Benefits of the Design Process for AI Product Development Practices} \label{sec:affordances of the process}
Several themes emerged about how participants viewed the Seamful XAI process can bring benefits to AI product development practices.

\subsubsection{Turning abstractions into specifics}
Participants expressed that the seamful XAI design process is focused, specific, and structured, \textbf{empowering them to transform abstract concepts in AI harms mitigation to specific instances.} Highlighting that the “process is tangible and generative” (P21), participants emphasized that it can be “effective in turning high-level guidelines into practice because seams are \textit{on-the-ground}” (P36). This, in turn, makes the process “more practically implementable with customers” (P38), elaborated by this participant:

\begin{quote}
Why the heck didn't we think like this before? I've used tons of AI Ethics tools. My clients hate how abstract they are. This [process] is specific and customers would love it...this is the next leap from impact or risk assessments.  (P03, a Product Manager)
\end{quote}

\noindent
In short, our process “takes the story multiple levels deeper into the solution space than just brainstorming about potential problems” (P39). The specificity, especially with breakdowns, can also help “kick-start the design process in the right way, which is very hard to do in practice” (P24). Participants appreciated the “hands-on nature of the process”, especially the “creative antagonistic thinking [they] did around breakdowns, and then getting [their] hands dirty to filter the seams'' (P06, P18). 
The hands-on nature can make abstract concepts tractable, elaborated by this participant:

\begin{quote}
Harm can be a nebulous word...This [process] grounds everything--
anchor on a specific breakdown, work through the [AI] lifecycle, and then use the seams...Unlike other RAI tools, the lifecycle stage allows you to pinpoint where things could go wrong (P25)
\end{quote}

\noindent
Moreover, our process has “a nice balance between top-down and bottom-up aspects: it has top-down ideas like going from breakdowns to seams with bottom-up aspects such as users generating the specific seams for a specific breakdown at a specific AI lifecycle” (P16). This is in contrast to “traditional impact assessments that are too top down” (P03).

\subsubsection{Fostering cross-disciplinary collaboration in a critically reflective manner}
Participants believed that the design process can facilitate \textbf{multi-stakeholder and interdisciplinary discussions}, providing an \textbf{inclusive} and \textbf{engaging way to lower the participation barrier.}

First, participants underscored how the process is an integrative approach that “stitches together all the layers of AI development in a way that can benefit all stakeholders” (P28). Even though some participants were aware of certain seams, they highlight a “lack of a framework to think through these issues” (P17), further elaborated by this participant:

\begin{quote}
This process gives end-to-end visibility, breaking silos rampant in tech companies. For example, if my PM foresees a seam in deployment, as a data scientist, I can address it upstream. It also helps with accountability; we know who is responsible for what. (P41)
\end{quote}

\noindent
Second, “by providing a shared medium and language to communicate” (P14), participants appreciated how the activity can empower inclusive multidisciplinary stakeholders to communicate effectively. They found the “vocabulary of seams helpful and terms like ‘breakdowns’ accessible” (P13). 
By “crafting the seam in plain words that anybody can understand, the process becomes very powerful. No one need[ed] a fancy machine learning degree to participate” (P40). This, in turn, can lessen feelings of “being a second-class citizen of a tech team” because ``on the [design] board, it's a level playing field. Be it a PM, a UXR, or a Designer...the process shows we’re all important” (P26).

Third, participants appreciated the inclusive “co-design nature of the activity” (P30). Envisioning it as “reflective tool for enacting Responsible AI in a step-by-step manner” (P06), participants described how the “process can challenge and break down [their] own assumptions” (P08):

\begin{quote}
I really like the co-design element. It made me feel included.
I can see doing this with my team. The flow was clear...
Having pre-filled examples of breakdowns and seams made it less intimidating and lowered the activation energy. (P34, a UX Researcher)
\end{quote}

\noindent
Lastly, participants strongly resonated with “how fun and engaging” the process was (P06) to lower the barriers for participation. Many “loved the simplicity and clarity”, emphasizing that it was “easier, more fun and hands-on than anticipated” (P12, P15). A key driver of engagement was the adversarial thinking participants did to anticipate seams by making the breakdowns happen: 

\begin{quote}
I had so much fun doing the super villain thinking! We're always expected to be the good guys...Getting a temporary license to be unshackled from that \textit{hero expectation} is \textit{freeing}--I don't have to hold back
...This is the X-factor of this process. 
This achieves the same goal without the heaviness or dullness of other RAI stuff.
(P24, emphasis added)
\end{quote}

\subsubsection{Taking a proactive design approach}
This design process promotes \textbf{a shift towards a proactive design approach in (X)AI}, which can mitigate the harms stemming from 
“the current reactive AI design--where we wait for things to fall apart from doing anything--slows down progress” and “is expensive in the long run” (P30, P22).  Participants found the process a “tractable proactive method that is a sharp departure from the reactive AI design” (P19). Beyond proactiveness, Seamful XAI carries the process till the end--”it goes beyond just asking 'what could go wrong?',\textit{ a point where most harms envisioning frameworks stop nowadays }“ (P30). 
By empowering people to leverage a seemingly negative seam into a positive, the process “transforms the information into actionable insights to make things right” (P30). This participant summarizes the value of a proactive process:

\begin{quote}
This proactive process lets us ask more meaningful questions...knowing the AI's blind spots can foster better feature engineering. It can inform where the AI may fall short and how we need to use seams to stop-gap the understanding. (P29, an XAI researcher)
\end{quote}
\noindent
A proactive approach, however, is not “the panacea of preventing AI harms” (P34):
\begin{quote}
Breakdowns are inevitable...This process can lessen the fallout. The proactive stance is like designing a car with a crumple zone that's lifesaving during crashes. Seamful information empowers users to handle AI mistakes without catastrophic failure. (P19)
\end{quote}

When asked if the process was too resource intensive, participants stressed that the “investment was worth it because you can pay a cost now or pay a higher cost later” (P30). 

\subsection{Challenges and Opportunities}
Every process has its challenges as well as areas of improvements. Below we touch on both.

Participants speculated two main challenges for practitioners to provide seamful XAI around \textbf{value tensions between teams} and \textbf{crafting a seam in a productive manner.} First, at an organizational level, \textit{competing incentives and value tensions between teams can impact which seams are shown or hidden.} “Each team has its own agenda; the product [team] has a go-to market strategy that may conflict with AI ethics division's-- who wins out?” (P34). \textit{To solve this problem}, participants suggested that “before teams even get started, they should put their cards on the table and get some alignment of goals...we don't want a fight when people are transferring seams to the reflection [card]” (P30). Participants, especially policy advisors, highlighted that “if we don't get alignment ahead of time, the process won't be as effective” (P28). They felt it's “imperative to harmonize business goals and the customer's interest” (P18) for the process to have the highest impact.

Second, at a process level, \textit{in the beginning, participants struggled to craft or frame the seam's information in an actionable manner to help the user albeit they were successful by the end of it.} 
While participants “got the concept of seams fairly easily”, “the challenge was how to phrase it such that it was actionable or helped the user” (P08, P12). Thankfully, \textbf{the scaffolding provided by the “crafting equation of expectations vs. reality” (P07) addressed this challenge.} “The equation was bang on; boiling it down to a recipe helped [them] devise the seams effectively” (P06).

Participants proposed \textit{three ways to improve the process} through (a) \textbf{knowing the impact of a seam}, (b) \textbf{adding action items}, and (c) \textbf{creating an organizationally accessible collection of seams.}
\textit{First, }they wished to know \textit{the impact of a particular seam on the outcome}. Knowing the impact would make the seamful information more actionable because it could “evaluate which seams to tackle first vs. later” (P23), a perspective effectively expressed here:

\begin{quote}
Show me what's the impact of this seam vs that seam: seam X impacts 35\% of the AI decision vs. seam Y impacts 10\% would be a killer feature because then I can prioritize using the impact data and the stickers on the board. (P04, an XAI focused UX researcher)
\end{quote}

\noindent
\textit{Second, }there was a demand \textit{for adding action items and recommendations along with the seamful information.} Participants suggested that the process would be “more powerful if [they] knew what to do with the seam” (P33). They recommended that adding the action item could “help less experienced end users to take the right action and make things consistent across the team” (P19). 

\textit{Third,} there was a popular suggestion to \textit{create a corpus of seams accessible across the organization.}
Having a corpus of seams categorized across “domains, problems, and AI life cycles stages would be a good starter pack” (P20). Participants had a few recommendations: one way could be “a design feature where users can flag useful seams, which are then added to the collection” (P01). Some highlighted an “inexpensive yet rich source of seams---the limitations documents” (P07). They argued that “turning the latent information in the limitations documentation in tech companies into activated knowledge can be a resourceful way to do seamful XAI” (P40).

\section{Discussion \& Implications}
Our study provides formative validation for the value of transferring and operationalizing the concept of seamful design into XAI. In this section, based on our empirical findings, we reflect on the roles Seamful XAI can play in expanding the design space of XAI, enabling RAI practices, and broader implications for research and practices around AI technologies.

\subsection{Expanding the Design Space of XAI}
 
A seamful outlook expands the XAI design space beyond the bounds of the algorithm. It extends the prior work in HCXAI by explicitly showing us how to use infrastructural mismatches (seams) in resourceful ways to augment explainability. Our findings showcase how seams can provide \textit{the peripheral vision of things outside the black-box} that are required to make complex decisions, can reveal the AI's blind spots, and can calibrate user trust in AI.
 
Seams bring to the forefront infrastructural mismatches that would otherwise likely be hidden or not considered. By doing so, as shown in Sec.~\ref{sec:enhancing explainability}, seamful information affords a unique type of understanding that helps with explainability -- \textit{knowing why the AI may not or cannot take certain things into account}. In complex real-world cases where the “ground truth is not black and white, we need to know \textit{when} we can trust the AI” (P30, emphasis added). If we know why the AI does well in one situation and not in another, we can calibrate when to rely on it:

\begin{quote}
Two important things happen: [first,] awareness of the broader context explains why the AI couldn't make the right call...[Second], it makes my trust on the AI nuanced. I know where it might fail but also where it'd succeed...It's no longer an all-or-nothing play. The AI is neither idiotic nor God-like (P32, an XAI researcher)
\vspace{-3pt}
\end{quote}

\noindent
\textit{A seamful outlook expands the epistemic canvas of XAI.} It “broadens the domain of things considered for XAI” (P33). This expansion has implications: 1)~it provides a unique lens of analysis  for XAI, 2)~it fosters “thinking of new interaction paradigms” (P28) that opportunistically leverage seams.

\subsection{Enabling resourceful, tractable, and actionable ways to do Responsible AI (RAI) } 
 
Our findings showcased how the seamful XAI design process enables us to enact RAI activities in a tractable and resourceful manner. The structured nature coupled with the on-the-ground situated nature of seams enable participants to transform abstract concepts like harms into specific instances. The Seamful XAI process extends the reach and impact of RAI activities in three ways. 
 
First, the process serves as a \textit{resourceful way of proactively mitigating harms}. There is a “two for the price of one” (P33) angle here: through the exercise of  anticipating and crafting seams, we have a ready-made list of actionable mismatches to work on, many of which can lead to harms. Thus, we have a proactive chance to neutralize the seams (rendering them irrelevant).
 
Second, our process \textit{not only generates a potential list of harms, but it also provides guidance on “what to do with them”} (P06). This guidance is facilitated by the filtering and designing with seams (going from step 2 to 3 in Fig.~\ref{fig:mural_design_process}). Participants found the exercise of filtering seams to be one of “the most innovative parts [because] it continued beyond merely generating harms, which is where most RAI approaches stop” (P04). Stakeholders can feel “overwhelmed and lost [when] they are left with a huge list and nothing to do about it” (P22). The filtering process not only forces us to critically think about how to leverage the seams but also prioritize the most important ones.
 
Third, seamful design encourages us to use the mismatches and leverage them in an opportunistic manner that empowers users~\cite{inmanribes2019,Chalmers2004,Chalmers2003}. In most “harm envisioning tasks, there is no impetus to leverage the harms” (P18). Thus, there may be missed opportunities where the user could have been empowered to act on the risks instead of just being made aware of them. For the seams that can be neutralized, our process allows that proactively. For the ones that we cannot do anything (e.g., model lags and organizational policy shifts), a seamful outlook prompts us to use uncertainties and limitations, as Gaver would put it, in a resourceful way~\cite{gaver2003ambiguity}.
\vspace{-5pt}

\subsection{Implications for Research \& Practice}
We share three implications of Seamful XAI on research and practice. \textit{First,} from an infrastructural view, a seamful approach allows us to leverage the imperfections in AI systems without needing to over-hype their capabilities or impose unachievable ideals. Understanding infrastructural mismatches can help users, especially non-AI experts, to be aware of the fallibility of AI systems. This awareness can shape how we think about trust calibration in AI, especially when addressing over-inflated expectations. Acknowledging AI is imperfect does not entail that we give up and do not work on problems~\cite{nilsson2016applying}. It simply means we are realistic about what we can solve and opportunistic about the rest. We leverage the gaps and use uncertainty as a resource instead of thinking of them as inherent flaws. 
An infrastructural view also changes how we design: instead of obsessing over perfection, we can be resourceful about Human-AI interaction designs that leverage inevitable imperfections to empower users with the tools to deal with the fallout from breakdowns.
 
\textit{Second,} a seamful XAI outlook can set the right expectations about the limits of our powers as technologists. Given that AI is a tricky design material~\cite{yang2020re,ehsan2023charting,AIdesignmaterial,DoveAIdesignmaterial} due to its stochasticity, changing, and non-deterministic nature, it is difficult to envision exactly how events unfold in use~\cite{phoebeseamless2022}. Given that breakdowns emerge in use, where developers or designers are unlikely to be around, it is in our best interests to empower end users \textit{with tools to address the fallout from breakdowns}. Moreover, a seamful outlook acknowledges that creators don't always have the control to dynamically change the system in real time (e.g., model updates may not be possible due to regulatory reasons and thus may lag organizational policy shifts). This realization can prompt changes in how we study AI failures and design fallback plans.
 
\textit{Third,} there is intrinsic value in adopting a proactive stance towards seamful XAI design \textit{even if we cannot anticipate or prevent all breakdowns}. Our proactive stance is a differentiating factor from the majority of prior work in seamful design where inventions exclusively dealt with seams post-deployment~\cite{BrollBenford2005,Chalmers2004,nilsson2016applying}. The core implication of the unbridgeable nature of the ``sociotechnical gap''~\cite{sociotechnicalgap2000}---the difference between what the system can technically afford and what users need---is that it may be impossible to exhaustively anticipate all seams. Despite that impossibility, there is value in a proactive stance because the very act of anticipatory exercises can (a) convert many “unknown unknowns” to “known unknowns” to promote proactive mitigation (discussed above) and (b) help us better understand and address the sociotechnical gap. 

No AI system is perfect. None will be. Seams are inevitable. A seamful view acknowledges the world as is. A seamless ideal projects a vision of the future, yet unrealized and arguably unrealizable. Seamfulness embraces the imperfect reality of spaces we inhabit and makes the most out of it.
  
\vspace{-5pt}
\section{Limitations \& Future Work}
We have taken a formative step on how to incorporate seamful design in the context of (X)AI. Given that this is a first step, the resulting insights should be scoped accordingly. We acknowledge limitations from using a scenario-based design, where there is data dependency on the particular context. Our findings should be interpreted as formative rather than evaluative. we hope that future work could evaluate how the process works longitudinally, explore how organizational resources impact engagement in it, and investigate the consequences of a seamful design approach in how it impacts the user experience. There might some cases (e.g., large-scale AI models) where proactive identification of all downstream breakdowns may not be possible \textit{before deployment}. In such cases, as shared in~\ref{subsec:step3_designing with seams}, the process should be used iteratively post-deployment.
\textit{For future iterations}, we are inspired by Phil Agre's notion of adopting a split identity, one where we simultaneously plant “one foot...in the craft work of design and the other foot in the reflexive work of critique”~\cite{agre_toward_1997}. We have planted one foot in design by introducing the notion of Seamful XAI. Now, we seek to learn from and with the broader HCI and XAI communities through critically constructive reflections.

\section{Conclusion}
In this paper, we demonstrated how seamful design advances Human-centered XAI. Through introducing and operationalizing the concept of Seamful XAI, we showed what seams look like in the context of XAI, and more broadly, AI. We developed and validated a design process to help stakeholders anticipate, filter, and design with seams. Our study showcased how the strategic revelation and concealment of sociotechnical and infrastructural seams enhance explainability, augment user agency, and, in fact, help identify and mitigate downstream AI harms. Recognizing that failures are a normal part of AI systems, seamful XAI embraces the incomplete and partial nature of AI systems, and instead of sweeping breakdowns under the rug, attempts to leverage seams as useful decision-making aids for end users. Through helping users live with systems that are inevitably imperfect, Seamful XAI aims to empower users to work with, through, and sometimes around such systems.

% \section{Acknowledgments}
\begin{acks}
With our deepest gratitude, we appreciate the time our participants generously invested in this project. This project would not have been possible without their involvement. Special thanks to Matthew Chalmers, one of the pioneers of Seamful Design, for engaging with us throughout this project and sharing constructive and kind feedback, which substantially helped with the conceptual development. We are grateful to members of the Fairness, Accountability, Transparency, and Ethics (FATE) group alongside other researchers and practitioners at Microsoft for their feedback that informed the empirical work. At multiple stages of the project over the years, we are appreciative of the generative discussions with Michael Muller, Sauvik Das, Sashank Varma, Munmun De Choudhury, Jennifer Wortman Vaughan, Kate Crawford, Mehrnoosh Sameki, Ben Noah, Gonzalo Ramos, and Mary Czerwinski. This project was partially supported by Microsoft Research and by the National Science Foundation under Grant No. 1928586.
\end{acks}
% \end{verbatim}
% so that the information contained therein can be more easily collected
% during the article metadata extraction phase, and to ensure
% consistency in the spelling of the section heading.

% Authors should not prepare this section as a numbered or unnumbered {\verb|\section|}; please use the ``{\verb|acks|}'' environment.

% \section{Appendices}
% If your work needs an appendix, add it before the
% ``\verb|\end{document}|'' command at the conclusion of your source
% document.

% Start the appendix with the ``\verb|appendix|'' command:
% \begin{verbatim}
%   \appendix
% \end{verbatim}
% and note that in the appendix, sections are lettered, not
% numbered. This document has two appendices, demonstrating the section
% and subsection identification method.

\bibliographystyle{ACM-Reference-Format}
\bibliography{sample-base}


\begin{thebibliography}{91}

%%% the Web address.

\ifx \showCODEN    \undefined \def \showCODEN     #1{\unskip}     \fi
\ifx \showDOI      \undefined \def \showDOI       #1{#1}\fi
\ifx \showISBNx    \undefined \def \showISBNx     #1{\unskip}     \fi
\ifx \showISBNxiii \undefined \def \showISBNxiii  #1{\unskip}     \fi
\ifx \showISSN     \undefined \def \showISSN      #1{\unskip}     \fi
\ifx \showLCCN     \undefined \def \showLCCN      #1{\unskip}     \fi
\ifx \shownote     \undefined \def \shownote      #1{#1}          \fi
\ifx \showarticletitle \undefined \def \showarticletitle #1{#1}   \fi
\ifx \showURL      \undefined \def \showURL       {\relax}        \fi
% invisible to TeX
\providecommand\bibfield[2]{#2}
\providecommand\bibinfo[2]{#2}
\providecommand\natexlab[1]{#1}
\providecommand\showeprint[2][]{arXiv:#2}

\bibitem[Ackerman(2000)]%
        {sociotechnicalgap2000}
\bibfield{author}{\bibinfo{person}{Mark~S. Ackerman}.} \bibinfo{year}{2000}\natexlab{}.
\newblock \showarticletitle{The Intellectual Challenge of CSCW: The Gap between Social Requirements and Technical Feasibility}.
\newblock \bibinfo{journal}{\emph{Hum.-Comput. Interact.}} \bibinfo{volume}{15}, \bibinfo{number}{2} (\bibinfo{date}{sep} \bibinfo{year}{2000}), \bibinfo{pages}{179–203}.
\newblock
\showISSN{0737-0024}
\urldef\tempurl%
\url{https://doi.org/10.1207/S15327051HCI1523_5}
\showDOI{\tempurl}


\bibitem[Agre(1997)]%
        {agre_toward_1997}
\bibfield{author}{\bibinfo{person}{P Agre}.} \bibinfo{year}{1997}\natexlab{}.
\newblock \showarticletitle{Toward a critical technical practice: {Lessons} learned in trying to reform {AI} in {Bowker}}.
\newblock \bibinfo{journal}{\emph{G., Star, S., Turner, W., and Gasser, L., eds, Social Science, Technical Systems and Cooperative Work: Beyond the Great Divide, Erlbaum}} (\bibinfo{year}{1997}).
\newblock


\bibitem[Aizenberg and Van Den~Hoven(2020)]%
        {aizenberg2020designing}
\bibfield{author}{\bibinfo{person}{Evgeni Aizenberg} {and} \bibinfo{person}{Jeroen Van Den~Hoven}.} \bibinfo{year}{2020}\natexlab{}.
\newblock \showarticletitle{Designing for human rights in AI}.
\newblock \bibinfo{journal}{\emph{Big Data \& Society}} \bibinfo{volume}{7}, \bibinfo{number}{2} (\bibinfo{year}{2020}), \bibinfo{pages}{2053951720949566}.
\newblock


\bibitem[Amershi(2020)]%
        {amershiplannigtofail}
\bibfield{author}{\bibinfo{person}{Saleema Amershi}.} \bibinfo{year}{2020}\natexlab{}.
\newblock \showarticletitle{Toward Responsible AI by Planning to Fail}. In \bibinfo{booktitle}{\emph{Proceedings of the 26th ACM SIGKDD International Conference on Knowledge Discovery and Data Mining}} (Virtual Event, CA, USA) \emph{(\bibinfo{series}{KDD '20})}. \bibinfo{publisher}{Association for Computing Machinery}, \bibinfo{address}{New York, NY, USA}, \bibinfo{pages}{3607}.
\newblock
\showISBNx{9781450379984}
\urldef\tempurl%
\url{https://doi.org/10.1145/3394486.3409557}
\showDOI{\tempurl}


\bibitem[Arrieta et~al\mbox{.}(2020)]%
        {arrieta2020explainable}
\bibfield{author}{\bibinfo{person}{Alejandro~Barredo Arrieta}, \bibinfo{person}{Natalia D{\'\i}az-Rodr{\'\i}guez}, \bibinfo{person}{Javier Del~Ser}, \bibinfo{person}{Adrien Bennetot}, \bibinfo{person}{Siham Tabik}, \bibinfo{person}{Alberto Barbado}, \bibinfo{person}{Salvador Garc{\'\i}a}, \bibinfo{person}{Sergio Gil-L{\'o}pez}, \bibinfo{person}{Daniel Molina}, \bibinfo{person}{Richard Benjamins}, {et~al\mbox{.}}} \bibinfo{year}{2020}\natexlab{}.
\newblock \showarticletitle{Explainable Artificial Intelligence (XAI): Concepts, taxonomies, opportunities and challenges toward responsible AI}.
\newblock \bibinfo{journal}{\emph{Information fusion}}  \bibinfo{volume}{58} (\bibinfo{year}{2020}), \bibinfo{pages}{82--115}.
\newblock


\bibitem[Bansal et~al\mbox{.}(2019)]%
        {BansalAccuracy}
\bibfield{author}{\bibinfo{person}{Gagan Bansal}, \bibinfo{person}{Besmira Nushi}, \bibinfo{person}{Ece Kamar}, \bibinfo{person}{Walter~S. Lasecki}, \bibinfo{person}{Daniel~S. Weld}, {and} \bibinfo{person}{Eric Horvitz}.} \bibinfo{year}{2019}\natexlab{}.
\newblock \showarticletitle{Beyond Accuracy: The Role of Mental Models in Human-AI Team Performance}.
\newblock \bibinfo{journal}{\emph{Proceedings of the AAAI Conference on Human Computation and Crowdsourcing}} \bibinfo{volume}{7}, \bibinfo{number}{1} (\bibinfo{date}{Oct.} \bibinfo{year}{2019}), \bibinfo{pages}{2--11}.
\newblock
\urldef\tempurl%
\url{https://doi.org/10.1609/hcomp.v7i1.5285}
\showDOI{\tempurl}


\bibitem[Bansal et~al\mbox{.}(2021)]%
        {bansal2021does}
\bibfield{author}{\bibinfo{person}{Gagan Bansal}, \bibinfo{person}{Tongshuang Wu}, \bibinfo{person}{Joyce Zhou}, \bibinfo{person}{Raymond Fok}, \bibinfo{person}{Besmira Nushi}, \bibinfo{person}{Ece Kamar}, \bibinfo{person}{Marco~Tulio Ribeiro}, {and} \bibinfo{person}{Daniel Weld}.} \bibinfo{year}{2021}\natexlab{}.
\newblock \showarticletitle{Does the whole exceed its parts? the effect of ai explanations on complementary team performance}. In \bibinfo{booktitle}{\emph{Proceedings of the 2021 CHI Conference on Human Factors in Computing Systems}}. \bibinfo{pages}{1--16}.
\newblock


\bibitem[Benjamins et~al\mbox{.}(2019)]%
        {benjamins2019responsible}
\bibfield{author}{\bibinfo{person}{Richard Benjamins}, \bibinfo{person}{Alberto Barbado}, {and} \bibinfo{person}{Daniel Sierra}.} \bibinfo{year}{2019}\natexlab{}.
\newblock \showarticletitle{Responsible AI by design in practice}.
\newblock \bibinfo{journal}{\emph{arXiv preprint arXiv:1909.12838}} (\bibinfo{year}{2019}).
\newblock


\bibitem[Brey(2012)]%
        {BreyAnticipatory}
\bibfield{author}{\bibinfo{person}{Philip~AE Brey}.} \bibinfo{year}{2012}\natexlab{}.
\newblock \showarticletitle{Anticipatory ethics for emerging technologies}.
\newblock \bibinfo{journal}{\emph{NanoEthics}} \bibinfo{volume}{6}, \bibinfo{number}{1} (\bibinfo{year}{2012}), \bibinfo{pages}{1--13}.
\newblock


\bibitem[Broll and Benford(2005)]%
        {BrollBenford2005}
\bibfield{author}{\bibinfo{person}{Gregor Broll} {and} \bibinfo{person}{Steve Benford}.} \bibinfo{year}{2005}\natexlab{}.
\newblock \showarticletitle{Seamful Design for Location-Based Mobile Games}. In \bibinfo{booktitle}{\emph{Entertainment Computing - ICEC 2005}}, \bibfield{editor}{\bibinfo{person}{Fumio Kishino}, \bibinfo{person}{Yoshifumi Kitamura}, \bibinfo{person}{Hirokazu Kato}, {and} \bibinfo{person}{Noriko Nagata}} (Eds.). \bibinfo{publisher}{Springer Berlin Heidelberg}, \bibinfo{address}{Berlin, Heidelberg}, \bibinfo{pages}{155--166}.
\newblock
\showISBNx{978-3-540-32054-8}


\bibitem[Chalmers(2003)]%
        {Chalmers2003seamful}
\bibfield{author}{\bibinfo{person}{Matthew Chalmers}.} \bibinfo{year}{2003}\natexlab{}.
\newblock \bibinfo{title}{Seamful Design and Ubicomp Infrastructure}.
\newblock
\newblock


\bibitem[Chalmers et~al\mbox{.}(2004)]%
        {Chalmers2004}
\bibfield{author}{\bibinfo{person}{Matthew Chalmers}, \bibinfo{person}{Andreas Dieberger}, \bibinfo{person}{Kristina Höök}, {and} \bibinfo{person}{Åsa Rudström}.} \bibinfo{year}{2004}\natexlab{}.
\newblock \showarticletitle{Social Navigation and Seamful Design}. In \bibinfo{booktitle}{\emph{In Japanese Journal of Cognitive Science, Special Issue on Social Navigation}}. \bibinfo{pages}{171--181}.
\newblock


\bibitem[Chalmers et~al\mbox{.}(2003)]%
        {Chalmers2003}
\bibfield{author}{\bibinfo{person}{M. Chalmers}, \bibinfo{person}{I. MacColl}, {and} \bibinfo{person}{M. Bell}.} \bibinfo{year}{2003}\natexlab{}.
\newblock \showarticletitle{Seamful design: showing the seams in wearable computing}. In \bibinfo{booktitle}{\emph{2003 IEE Eurowearable}}. \bibinfo{pages}{11--16}.
\newblock
\urldef\tempurl%
\url{https://doi.org/10.1049/ic:20030140}
\showDOI{\tempurl}


\bibitem[Charmaz(2014)]%
        {Charmaz2014}
\bibfield{author}{\bibinfo{person}{Kathy Charmaz}.} \bibinfo{year}{2014}\natexlab{}.
\newblock \bibinfo{booktitle}{\emph{{Constructing Grounded Theory (Introducing Qualitative Methods series) 2nd Edition}}}.
\newblock \bibinfo{publisher}{Sage}, \bibinfo{address}{London}.
\newblock


\bibitem[Chybowski et~al\mbox{.}(2018)]%
        {chybowski2018applying}
\bibfield{author}{\bibinfo{person}{Leszek Chybowski}, \bibinfo{person}{Katarzyna Gawdzi{\'n}ska}, {and} \bibinfo{person}{Valeri Souchkov}.} \bibinfo{year}{2018}\natexlab{}.
\newblock \showarticletitle{Applying the anticipatory failure determination at a very early stage of a system’s development: overview and case study}.
\newblock \bibinfo{journal}{\emph{Multidisciplinary Aspects of Production Engineering}}  \bibinfo{volume}{1} (\bibinfo{year}{2018}).
\newblock


\bibitem[Delgado et~al\mbox{.}(2021)]%
        {delgado2021stakeholder}
\bibfield{author}{\bibinfo{person}{Fernando Delgado}, \bibinfo{person}{Stephen Yang}, \bibinfo{person}{Michael Madaio}, {and} \bibinfo{person}{Qian Yang}.} \bibinfo{year}{2021}\natexlab{}.
\newblock \showarticletitle{Stakeholder Participation in AI: Beyond" Add Diverse Stakeholders and Stir"}.
\newblock \bibinfo{journal}{\emph{arXiv preprint arXiv:2111.01122}} (\bibinfo{year}{2021}).
\newblock


\bibitem[Deng et~al\mbox{.}(2022)]%
        {deng2022exploring}
\bibfield{author}{\bibinfo{person}{Wesley~Hanwen Deng}, \bibinfo{person}{Manish Nagireddy}, \bibinfo{person}{Michelle Seng~Ah Lee}, \bibinfo{person}{Jatinder Singh}, \bibinfo{person}{Zhiwei~Steven Wu}, \bibinfo{person}{Kenneth Holstein}, {and} \bibinfo{person}{Haiyi Zhu}.} \bibinfo{year}{2022}\natexlab{}.
\newblock \showarticletitle{Exploring How Machine Learning Practitioners (Try To) Use Fairness Toolkits}.
\newblock \bibinfo{journal}{\emph{arXiv preprint arXiv:2205.06922}} (\bibinfo{year}{2022}).
\newblock


\bibitem[Dhanorkar et~al\mbox{.}(2021)]%
        {dhanorkar2021needs}
\bibfield{author}{\bibinfo{person}{Shipi Dhanorkar}, \bibinfo{person}{Christine~T Wolf}, \bibinfo{person}{Kun Qian}, \bibinfo{person}{Anbang Xu}, \bibinfo{person}{Lucian Popa}, {and} \bibinfo{person}{Yunyao Li}.} \bibinfo{year}{2021}\natexlab{}.
\newblock \showarticletitle{Who needs to know what, when?: Broadening the Explainable AI (XAI) Design Space by Looking at Explanations Across the AI Lifecycle}. In \bibinfo{booktitle}{\emph{Designing Interactive Systems Conference 2021}}. \bibinfo{pages}{1591--1602}.
\newblock


\bibitem[Dodge et~al\mbox{.}(2019)]%
        {dodge2019explaining}
\bibfield{author}{\bibinfo{person}{Jonathan Dodge}, \bibinfo{person}{Q~Vera Liao}, \bibinfo{person}{Yunfeng Zhang}, \bibinfo{person}{Rachel~KE Bellamy}, {and} \bibinfo{person}{Casey Dugan}.} \bibinfo{year}{2019}\natexlab{}.
\newblock \showarticletitle{Explaining models: an empirical study of how explanations impact fairness judgment}. In \bibinfo{booktitle}{\emph{Proceedings of the 24th international conference on intelligent user interfaces}}. \bibinfo{pages}{275--285}.
\newblock


\bibitem[Dove et~al\mbox{.}(2017)]%
        {DoveAIdesignmaterial}
\bibfield{author}{\bibinfo{person}{Graham Dove}, \bibinfo{person}{Kim Halskov}, \bibinfo{person}{Jodi Forlizzi}, {and} \bibinfo{person}{John Zimmerman}.} \bibinfo{year}{2017}\natexlab{}.
\newblock \showarticletitle{UX Design Innovation: Challenges for Working with Machine Learning as a Design Material}. In \bibinfo{booktitle}{\emph{Proceedings of the 2017 CHI Conference on Human Factors in Computing Systems}} (Denver, Colorado, USA) \emph{(\bibinfo{series}{CHI '17})}. \bibinfo{publisher}{Association for Computing Machinery}, \bibinfo{address}{New York, NY, USA}, \bibinfo{pages}{278–288}.
\newblock
\showISBNx{9781450346559}
\urldef\tempurl%
\url{https://doi.org/10.1145/3025453.3025739}
\showDOI{\tempurl}


\bibitem[Ehsan et~al\mbox{.}(2021a)]%
        {ehsan2021expanding}
\bibfield{author}{\bibinfo{person}{Upol Ehsan}, \bibinfo{person}{Q~Vera Liao}, \bibinfo{person}{Michael Muller}, \bibinfo{person}{Mark~O Riedl}, {and} \bibinfo{person}{Justin~D Weisz}.} \bibinfo{year}{2021}\natexlab{a}.
\newblock \showarticletitle{Expanding explainability: Towards social transparency in ai systems}. In \bibinfo{booktitle}{\emph{Proceedings of the 2021 CHI Conference on Human Factors in Computing Systems}}. \bibinfo{pages}{1--19}.
\newblock


\bibitem[Ehsan and Riedl(2020)]%
        {ehsan2020human}
\bibfield{author}{\bibinfo{person}{Upol Ehsan} {and} \bibinfo{person}{Mark~O Riedl}.} \bibinfo{year}{2020}\natexlab{}.
\newblock \showarticletitle{Human-centered explainable ai: Towards a reflective sociotechnical approach}. In \bibinfo{booktitle}{\emph{International Conference on Human-Computer Interaction}}. Springer, \bibinfo{pages}{449--466}.
\newblock


\bibitem[Ehsan and Riedl(2021)]%
        {ehsan2021explainability}
\bibfield{author}{\bibinfo{person}{Upol Ehsan} {and} \bibinfo{person}{Mark~O Riedl}.} \bibinfo{year}{2021}\natexlab{}.
\newblock \showarticletitle{Explainability pitfalls: Beyond dark patterns in explainable AI}.
\newblock \bibinfo{journal}{\emph{arXiv preprint arXiv:2109.12480}} (\bibinfo{year}{2021}).
\newblock


\bibitem[Ehsan et~al\mbox{.}(2023)]%
        {ehsan2023charting}
\bibfield{author}{\bibinfo{person}{Upol Ehsan}, \bibinfo{person}{Koustuv Saha}, \bibinfo{person}{Munmun De~Choudhury}, {and} \bibinfo{person}{Mark~O Riedl}.} \bibinfo{year}{2023}\natexlab{}.
\newblock \showarticletitle{Charting the Sociotechnical Gap in Explainable AI: A Framework to Address the Gap in XAI}.
\newblock \bibinfo{journal}{\emph{Proceedings of the ACM on Human-Computer Interaction}} \bibinfo{volume}{7}, \bibinfo{number}{CSCW1} (\bibinfo{year}{2023}), \bibinfo{pages}{1--32}.
\newblock


\bibitem[Ehsan et~al\mbox{.}(2022)]%
        {ehsan2022algorithmic}
\bibfield{author}{\bibinfo{person}{Upol Ehsan}, \bibinfo{person}{Ranjit Singh}, \bibinfo{person}{Jacob Metcalf}, {and} \bibinfo{person}{Mark Riedl}.} \bibinfo{year}{2022}\natexlab{}.
\newblock \showarticletitle{The Algorithmic Imprint}. In \bibinfo{booktitle}{\emph{2022 ACM Conference on Fairness, Accountability, and Transparency}}. \bibinfo{pages}{1305--1317}.
\newblock


\bibitem[Ehsan et~al\mbox{.}(2021b)]%
        {ehsan2021operationalizing}
\bibfield{author}{\bibinfo{person}{Upol Ehsan}, \bibinfo{person}{Philipp Wintersberger}, \bibinfo{person}{Q~Vera Liao}, \bibinfo{person}{Martina Mara}, \bibinfo{person}{Marc Streit}, \bibinfo{person}{Sandra Wachter}, \bibinfo{person}{Andreas Riener}, {and} \bibinfo{person}{Mark~O Riedl}.} \bibinfo{year}{2021}\natexlab{b}.
\newblock \showarticletitle{Operationalizing human-centered perspectives in explainable AI}. In \bibinfo{booktitle}{\emph{Extended Abstracts of the 2021 CHI Conference on Human Factors in Computing Systems}}. \bibinfo{pages}{1--6}.
\newblock


\bibitem[Gaver et~al\mbox{.}(2003)]%
        {gaver2003ambiguity}
\bibfield{author}{\bibinfo{person}{William~W Gaver}, \bibinfo{person}{Jacob Beaver}, {and} \bibinfo{person}{Steve Benford}.} \bibinfo{year}{2003}\natexlab{}.
\newblock \showarticletitle{Ambiguity as a resource for design}. In \bibinfo{booktitle}{\emph{Proceedings of the SIGCHI conference on Human factors in computing systems}}. \bibinfo{pages}{233--240}.
\newblock


\bibitem[Gebru et~al\mbox{.}(2021)]%
        {gebru2021datasheets}
\bibfield{author}{\bibinfo{person}{Timnit Gebru}, \bibinfo{person}{Jamie Morgenstern}, \bibinfo{person}{Briana Vecchione}, \bibinfo{person}{Jennifer~Wortman Vaughan}, \bibinfo{person}{Hanna Wallach}, \bibinfo{person}{Hal~Daum{\'e} Iii}, {and} \bibinfo{person}{Kate Crawford}.} \bibinfo{year}{2021}\natexlab{}.
\newblock \showarticletitle{Datasheets for datasets}.
\newblock \bibinfo{journal}{\emph{Commun. ACM}} \bibinfo{volume}{64}, \bibinfo{number}{12} (\bibinfo{year}{2021}), \bibinfo{pages}{86--92}.
\newblock


\bibitem[Ghai et~al\mbox{.}(2021)]%
        {ghai2021explainable}
\bibfield{author}{\bibinfo{person}{Bhavya Ghai}, \bibinfo{person}{Q~Vera Liao}, \bibinfo{person}{Yunfeng Zhang}, \bibinfo{person}{Rachel Bellamy}, {and} \bibinfo{person}{Klaus Mueller}.} \bibinfo{year}{2021}\natexlab{}.
\newblock \showarticletitle{Explainable active learning (xal) toward ai explanations as interfaces for machine teachers}.
\newblock \bibinfo{journal}{\emph{Proceedings of the ACM on Human-Computer Interaction}} \bibinfo{volume}{4}, \bibinfo{number}{CSCW3} (\bibinfo{year}{2021}), \bibinfo{pages}{1--28}.
\newblock


\bibitem[Gilpin et~al\mbox{.}(2018)]%
        {gilpin2018explaining}
\bibfield{author}{\bibinfo{person}{Leilani~H Gilpin}, \bibinfo{person}{David Bau}, \bibinfo{person}{Ben~Z Yuan}, \bibinfo{person}{Ayesha Bajwa}, \bibinfo{person}{Michael Specter}, {and} \bibinfo{person}{Lalana Kagal}.} \bibinfo{year}{2018}\natexlab{}.
\newblock \showarticletitle{Explaining explanations: An overview of interpretability of machine learning}. In \bibinfo{booktitle}{\emph{2018 IEEE 5th International Conference on data science and advanced analytics (DSAA)}}. IEEE, \bibinfo{pages}{80--89}.
\newblock


\bibitem[Gonz{\'a}lez et~al\mbox{.}(2021)]%
        {GonzalezExplanations}
\bibfield{author}{\bibinfo{person}{Ana~Valeria Gonz{\'a}lez}, \bibinfo{person}{Gagan Bansal}, \bibinfo{person}{Angela Fan}, \bibinfo{person}{Yashar Mehdad}, \bibinfo{person}{Robin Jia}, {and} \bibinfo{person}{Srinivasan Iyer}.} \bibinfo{year}{2021}\natexlab{}.
\newblock \showarticletitle{Do Explanations Help Users Detect Errors in Open-Domain {QA}? An Evaluation of Spoken vs. Visual Explanations}. In \bibinfo{booktitle}{\emph{Findings of the Association for Computational Linguistics: ACL-IJCNLP 2021}}. \bibinfo{publisher}{Association for Computational Linguistics}, \bibinfo{address}{Online}, \bibinfo{pages}{1103--1116}.
\newblock
\urldef\tempurl%
\url{https://doi.org/10.18653/v1/2021.findings-acl.95}
\showDOI{\tempurl}


\bibitem[Guidotti et~al\mbox{.}(2018)]%
        {guidotti2018survey}
\bibfield{author}{\bibinfo{person}{Riccardo Guidotti}, \bibinfo{person}{Anna Monreale}, \bibinfo{person}{Salvatore Ruggieri}, \bibinfo{person}{Franco Turini}, \bibinfo{person}{Fosca Giannotti}, {and} \bibinfo{person}{Dino Pedreschi}.} \bibinfo{year}{2018}\natexlab{}.
\newblock \showarticletitle{A survey of methods for explaining black box models}.
\newblock \bibinfo{journal}{\emph{ACM computing surveys (CSUR)}} \bibinfo{volume}{51}, \bibinfo{number}{5} (\bibinfo{year}{2018}), \bibinfo{pages}{1--42}.
\newblock


\bibitem[Gunning et~al\mbox{.}(2019)]%
        {gunning2019xai}
\bibfield{author}{\bibinfo{person}{David Gunning}, \bibinfo{person}{Mark Stefik}, \bibinfo{person}{Jaesik Choi}, \bibinfo{person}{Timothy Miller}, \bibinfo{person}{Simone Stumpf}, {and} \bibinfo{person}{Guang-Zhong Yang}.} \bibinfo{year}{2019}\natexlab{}.
\newblock \showarticletitle{XAI—Explainable artificial intelligence}.
\newblock \bibinfo{journal}{\emph{Science Robotics}} \bibinfo{volume}{4}, \bibinfo{number}{37} (\bibinfo{year}{2019}).
\newblock


\bibitem[Halfaker and Geiger(2020)]%
        {halfaker2020ores}
\bibfield{author}{\bibinfo{person}{Aaron Halfaker} {and} \bibinfo{person}{R~Stuart Geiger}.} \bibinfo{year}{2020}\natexlab{}.
\newblock \showarticletitle{Ores: Lowering barriers with participatory machine learning in wikipedia}.
\newblock \bibinfo{journal}{\emph{Proceedings of the ACM on Human-Computer Interaction}} \bibinfo{volume}{4}, \bibinfo{number}{CSCW2} (\bibinfo{year}{2020}), \bibinfo{pages}{1--37}.
\newblock


\bibitem[Heger et~al\mbox{.}(2022)]%
        {amy2022}
\bibfield{author}{\bibinfo{person}{Amy Heger}, \bibinfo{person}{Liz~B. Marquis}, \bibinfo{person}{Mihaela Vorvoreanu}, \bibinfo{person}{Hanna Wallach}, {and} \bibinfo{person}{Jennifer~W. Vaughan}.} \bibinfo{year}{2022}\natexlab{}.
\newblock \showarticletitle{Understanding Machine Learning Practitioners’ Data Documentation Perceptions, Needs, Challenges, and Desiderata}.
\newblock \bibinfo{journal}{\emph{Proceedings of the ACM on Human-Computer Interaction}} \bibinfo{volume}{6}, \bibinfo{number}{CSCW2} (\bibinfo{year}{2022}), \bibinfo{pages}{1--30}.
\newblock


\bibitem[Hengesbach(2022)]%
        {seamfulviz2022}
\bibfield{author}{\bibinfo{person}{Nicole Hengesbach}.} \bibinfo{year}{2022}\natexlab{}.
\newblock \showarticletitle{Undoing Seamlessness: Exploring Seams for Critical Visualization}. In \bibinfo{booktitle}{\emph{Extended Abstracts of the 2022 CHI Conference on Human Factors in Computing Systems}} (New Orleans, LA, USA) \emph{(\bibinfo{series}{CHI EA '22})}. \bibinfo{publisher}{Association for Computing Machinery}, \bibinfo{address}{New York, NY, USA}, Article \bibinfo{articleno}{364}, \bibinfo{numpages}{7}~pages.
\newblock
\showISBNx{9781450391566}
\urldef\tempurl%
\url{https://doi.org/10.1145/3491101.3519703}
\showDOI{\tempurl}


\bibitem[Holmquist(2017)]%
        {AIdesignmaterial}
\bibfield{author}{\bibinfo{person}{Lars~Erik Holmquist}.} \bibinfo{year}{2017}\natexlab{}.
\newblock \showarticletitle{Intelligence on Tap: Artificial Intelligence as a New Design Material}.
\newblock \bibinfo{journal}{\emph{Interactions}} \bibinfo{volume}{24}, \bibinfo{number}{4} (\bibinfo{date}{jun} \bibinfo{year}{2017}), \bibinfo{pages}{28–33}.
\newblock
\showISSN{1072-5520}
\urldef\tempurl%
\url{https://doi.org/10.1145/3085571}
\showDOI{\tempurl}


\bibitem[Holstein et~al\mbox{.}(2019)]%
        {holstein2019improving}
\bibfield{author}{\bibinfo{person}{Kenneth Holstein}, \bibinfo{person}{Jennifer Wortman~Vaughan}, \bibinfo{person}{Hal Daum{\'e}~III}, \bibinfo{person}{Miro Dudik}, {and} \bibinfo{person}{Hanna Wallach}.} \bibinfo{year}{2019}\natexlab{}.
\newblock \showarticletitle{Improving fairness in machine learning systems: What do industry practitioners need?}. In \bibinfo{booktitle}{\emph{Proceedings of the 2019 CHI conference on human factors in computing systems}}. \bibinfo{pages}{1--16}.
\newblock


\bibitem[Hong et~al\mbox{.}(2021)]%
        {hong2021planning}
\bibfield{author}{\bibinfo{person}{Matthew~K Hong}, \bibinfo{person}{Adam Fourney}, \bibinfo{person}{Derek DeBellis}, {and} \bibinfo{person}{Saleema Amershi}.} \bibinfo{year}{2021}\natexlab{}.
\newblock \showarticletitle{Planning for natural language failures with the ai playbook}. In \bibinfo{booktitle}{\emph{Proceedings of the 2021 CHI Conference on Human Factors in Computing Systems}}. \bibinfo{pages}{1--11}.
\newblock


\bibitem[Hong et~al\mbox{.}(2020)]%
        {hong2020human}
\bibfield{author}{\bibinfo{person}{Sungsoo~Ray Hong}, \bibinfo{person}{Jessica Hullman}, {and} \bibinfo{person}{Enrico Bertini}.} \bibinfo{year}{2020}\natexlab{}.
\newblock \showarticletitle{Human factors in model interpretability: Industry practices, challenges, and needs}.
\newblock \bibinfo{journal}{\emph{Proceedings of the ACM on Human-Computer Interaction}} \bibinfo{volume}{4}, \bibinfo{number}{CSCW1} (\bibinfo{year}{2020}), \bibinfo{pages}{1--26}.
\newblock


\bibitem[Ilevbare et~al\mbox{.}(2013)]%
        {ilevbare2013review}
\bibfield{author}{\bibinfo{person}{Imoh~M Ilevbare}, \bibinfo{person}{David Probert}, {and} \bibinfo{person}{Robert Phaal}.} \bibinfo{year}{2013}\natexlab{}.
\newblock \showarticletitle{A review of TRIZ, and its benefits and challenges in practice}.
\newblock \bibinfo{journal}{\emph{Technovation}} \bibinfo{volume}{33}, \bibinfo{number}{2-3} (\bibinfo{year}{2013}), \bibinfo{pages}{30--37}.
\newblock


\bibitem[Inman and Ribes(2019)]%
        {inmanribes2019}
\bibfield{author}{\bibinfo{person}{Sarah Inman} {and} \bibinfo{person}{David Ribes}.} \bibinfo{year}{2019}\natexlab{}.
\newblock \showarticletitle{"Beautiful Seams": Strategic Revelations and Concealments}. In \bibinfo{booktitle}{\emph{Proceedings of the 2019 CHI Conference on Human Factors in Computing Systems}} (Glasgow, Scotland Uk) \emph{(\bibinfo{series}{CHI '19})}. \bibinfo{publisher}{Association for Computing Machinery}, \bibinfo{address}{New York, NY, USA}, \bibinfo{pages}{1–14}.
\newblock
\showISBNx{9781450359702}
\urldef\tempurl%
\url{https://doi.org/10.1145/3290605.3300508}
\showDOI{\tempurl}


\bibitem[Jacobs et~al\mbox{.}(2021)]%
        {jacobs2021designing}
\bibfield{author}{\bibinfo{person}{Maia Jacobs}, \bibinfo{person}{Jeffrey He}, \bibinfo{person}{Melanie F.~Pradier}, \bibinfo{person}{Barbara Lam}, \bibinfo{person}{Andrew~C Ahn}, \bibinfo{person}{Thomas~H McCoy}, \bibinfo{person}{Roy~H Perlis}, \bibinfo{person}{Finale Doshi-Velez}, {and} \bibinfo{person}{Krzysztof~Z Gajos}.} \bibinfo{year}{2021}\natexlab{}.
\newblock \showarticletitle{Designing AI for trust and collaboration in time-constrained medical decisions: a sociotechnical lens}. In \bibinfo{booktitle}{\emph{Proceedings of the 2021 CHI Conference on Human Factors in Computing Systems}}. \bibinfo{pages}{1--14}.
\newblock


\bibitem[Karimi et~al\mbox{.}(2021)]%
        {karimi2021algorithmic}
\bibfield{author}{\bibinfo{person}{Amir-Hossein Karimi}, \bibinfo{person}{Bernhard Sch{\"o}lkopf}, {and} \bibinfo{person}{Isabel Valera}.} \bibinfo{year}{2021}\natexlab{}.
\newblock \showarticletitle{Algorithmic recourse: from counterfactual explanations to interventions}. In \bibinfo{booktitle}{\emph{Proceedings of the 2021 ACM conference on fairness, accountability, and transparency}}. \bibinfo{pages}{353--362}.
\newblock


\bibitem[Kashfi et~al\mbox{.}(2017)]%
        {Kashfi2017}
\bibfield{author}{\bibinfo{person}{Pariya Kashfi}, \bibinfo{person}{Agneta Nilsson}, {and} \bibinfo{person}{Robert Feldt}.} \bibinfo{year}{2017}\natexlab{}.
\newblock \showarticletitle{Integrating User eXperience practices into software development processes: implications of the UX characteristics}.
\newblock \bibinfo{journal}{\emph{PeerJ Computer Science}}  \bibinfo{volume}{3} (\bibinfo{date}{10} \bibinfo{year}{2017}), \bibinfo{pages}{e130}.
\newblock
\urldef\tempurl%
\url{https://doi.org/10.7717/peerj-cs.130}
\showDOI{\tempurl}


\bibitem[Kaur et~al\mbox{.}(2022)]%
        {kaur2022sensible}
\bibfield{author}{\bibinfo{person}{Harmanpreet Kaur}, \bibinfo{person}{Eytan Adar}, \bibinfo{person}{Eric Gilbert}, {and} \bibinfo{person}{Cliff Lampe}.} \bibinfo{year}{2022}\natexlab{}.
\newblock \showarticletitle{Sensible AI: Re-imagining Interpretability and Explainability using Sensemaking Theory}.
\newblock \bibinfo{journal}{\emph{arXiv preprint arXiv:2205.05057}} (\bibinfo{year}{2022}).
\newblock


\bibitem[Kayacik et~al\mbox{.}(2019)]%
        {Kayacik2019}
\bibfield{author}{\bibinfo{person}{Claire Kayacik}, \bibinfo{person}{Sherol Chen}, \bibinfo{person}{Signe Noerly}, \bibinfo{person}{Jess Holbrook}, \bibinfo{person}{Adam Roberts}, {and} \bibinfo{person}{Douglas Eck}.} \bibinfo{year}{2019}\natexlab{}.
\newblock \showarticletitle{Identifying the Intersections: User Experience + Research Scientist Collaboration in a Generative Machine Learning Interface}. In \bibinfo{booktitle}{\emph{Extended Abstracts of the 2019 CHI Conference on Human Factors in Computing Systems}} (Glasgow, Scotland Uk) \emph{(\bibinfo{series}{CHI EA '19})}. \bibinfo{publisher}{Association for Computing Machinery}, \bibinfo{address}{New York, NY, USA}, \bibinfo{pages}{1–8}.
\newblock
\showISBNx{9781450359719}
\urldef\tempurl%
\url{https://doi.org/10.1145/3290607.3299059}
\showDOI{\tempurl}


\bibitem[Keane and Smyth(2020)]%
        {keane_good_2020}
\bibfield{author}{\bibinfo{person}{Mark~T. Keane} {and} \bibinfo{person}{Barry Smyth}.} \bibinfo{year}{2020}\natexlab{}.
\newblock \showarticletitle{Good {Counterfactuals} and {Where} to {Find} {Them}: {A} {Case}-{Based} {Technique} for {Generating} {Counterfactuals} for {Explainable} {AI} ({XAI})}. In \bibinfo{booktitle}{\emph{Case-{Based} {Reasoning} {Research} and {Development}}} \emph{(\bibinfo{series}{Lecture {Notes} in {Computer} {Science}})}, \bibfield{editor}{\bibinfo{person}{Ian Watson} {and} \bibinfo{person}{Rosina Weber}} (Eds.). \bibinfo{publisher}{Springer International Publishing}, \bibinfo{address}{Cham}, \bibinfo{pages}{163--178}.
\newblock
\showISBNx{978-3-030-58342-2}
\urldef\tempurl%
\url{https://doi.org/10.1007/978-3-030-58342-2_11}
\showDOI{\tempurl}


\bibitem[Leonardi(2013)]%
        {leonardi_theoretical_2013}
\bibfield{author}{\bibinfo{person}{Paul~M Leonardi}.} \bibinfo{year}{2013}\natexlab{}.
\newblock \showarticletitle{Theoretical foundations for the study of sociomateriality}.
\newblock \bibinfo{journal}{\emph{Information and organization}} \bibinfo{volume}{23}, \bibinfo{number}{2} (\bibinfo{year}{2013}), \bibinfo{pages}{59--76}.
\newblock
\newblock
\shownote{Publisher: Elsevier}.


\bibitem[Liao et~al\mbox{.}(2020)]%
        {liao2020questioning}
\bibfield{author}{\bibinfo{person}{Q~Vera Liao}, \bibinfo{person}{Daniel Gruen}, {and} \bibinfo{person}{Sarah Miller}.} \bibinfo{year}{2020}\natexlab{}.
\newblock \showarticletitle{Questioning the AI: informing design practices for explainable AI user experiences}. In \bibinfo{booktitle}{\emph{Proceedings of the 2020 CHI Conference on Human Factors in Computing Systems}}. \bibinfo{pages}{1--15}.
\newblock


\bibitem[Liao and Varshney(2021)]%
        {liao2021human}
\bibfield{author}{\bibinfo{person}{Q~Vera Liao} {and} \bibinfo{person}{Kush~R Varshney}.} \bibinfo{year}{2021}\natexlab{}.
\newblock \showarticletitle{Human-Centered Explainable AI (XAI): From Algorithms to User Experiences}.
\newblock \bibinfo{journal}{\emph{arXiv preprint arXiv:2110.10790}} (\bibinfo{year}{2021}).
\newblock


\bibitem[Liao et~al\mbox{.}(2022)]%
        {liao2022connecting}
\bibfield{author}{\bibinfo{person}{Q~Vera Liao}, \bibinfo{person}{Yunfeng Zhang}, \bibinfo{person}{Ronny Luss}, \bibinfo{person}{Finale Doshi-Velez}, {and} \bibinfo{person}{Amit Dhurandhar}.} \bibinfo{year}{2022}\natexlab{}.
\newblock \showarticletitle{Connecting Algorithmic Research and Usage Contexts: A Perspective of Contextualized Evaluation for Explainable AI}.
\newblock \bibinfo{journal}{\emph{arXiv preprint arXiv:2206.10847}} (\bibinfo{year}{2022}).
\newblock


\bibitem[Lipton(2016)]%
        {lipton2016mythos}
\bibfield{author}{\bibinfo{person}{Zachary~C Lipton}.} \bibinfo{year}{2016}\natexlab{}.
\newblock \showarticletitle{The mythos of model interpretability}.
\newblock \bibinfo{journal}{\emph{arXiv preprint arXiv:1606.03490}} (\bibinfo{year}{2016}).
\newblock


\bibitem[Lyons et~al\mbox{.}(2021)]%
        {lyons2021conceptualising}
\bibfield{author}{\bibinfo{person}{Henrietta Lyons}, \bibinfo{person}{Eduardo Velloso}, {and} \bibinfo{person}{Tim Miller}.} \bibinfo{year}{2021}\natexlab{}.
\newblock \showarticletitle{Conceptualising contestability: Perspectives on contesting algorithmic decisions}.
\newblock \bibinfo{journal}{\emph{Proceedings of the ACM on Human-Computer Interaction}} \bibinfo{volume}{5}, \bibinfo{number}{CSCW1} (\bibinfo{year}{2021}), \bibinfo{pages}{1--25}.
\newblock


\bibitem[Madaio et~al\mbox{.}(2022)]%
        {madaio2022assessing}
\bibfield{author}{\bibinfo{person}{Michael Madaio}, \bibinfo{person}{Lisa Egede}, \bibinfo{person}{Hariharan Subramonyam}, \bibinfo{person}{Jennifer Wortman~Vaughan}, {and} \bibinfo{person}{Hanna Wallach}.} \bibinfo{year}{2022}\natexlab{}.
\newblock \showarticletitle{Assessing the Fairness of AI Systems: AI Practitioners' Processes, Challenges, and Needs for Support}.
\newblock \bibinfo{journal}{\emph{Proceedings of the ACM on Human-Computer Interaction}} \bibinfo{volume}{6}, \bibinfo{number}{CSCW1} (\bibinfo{year}{2022}).
\newblock


\bibitem[Madaio et~al\mbox{.}(2020a)]%
        {Madaio2020}
\bibfield{author}{\bibinfo{person}{Michael~A. Madaio}, \bibinfo{person}{Luke Stark}, \bibinfo{person}{Jennifer Wortman~Vaughan}, {and} \bibinfo{person}{Hanna Wallach}.} \bibinfo{year}{2020}\natexlab{a}.
\newblock \showarticletitle{Co-Designing Checklists to Understand Organizational Challenges and Opportunities around Fairness in AI}. In \bibinfo{booktitle}{\emph{Proceedings of the 2020 CHI Conference on Human Factors in Computing Systems}} (Honolulu, HI, USA) \emph{(\bibinfo{series}{CHI '20})}. \bibinfo{publisher}{Association for Computing Machinery}, \bibinfo{address}{New York, NY, USA}, \bibinfo{pages}{1–14}.
\newblock
\showISBNx{9781450367080}
\urldef\tempurl%
\url{https://doi.org/10.1145/3313831.3376445}
\showDOI{\tempurl}


\bibitem[Madaio et~al\mbox{.}(2020b)]%
        {madaio2020co}
\bibfield{author}{\bibinfo{person}{Michael~A Madaio}, \bibinfo{person}{Luke Stark}, \bibinfo{person}{Jennifer Wortman~Vaughan}, {and} \bibinfo{person}{Hanna Wallach}.} \bibinfo{year}{2020}\natexlab{b}.
\newblock \showarticletitle{Co-designing checklists to understand organizational challenges and opportunities around fairness in AI}. In \bibinfo{booktitle}{\emph{Proceedings of the 2020 CHI Conference on Human Factors in Computing Systems}}. \bibinfo{pages}{1--14}.
\newblock


\bibitem[Metcalf et~al\mbox{.}(2021)]%
        {metcalfimpact2021}
\bibfield{author}{\bibinfo{person}{Jacob Metcalf}, \bibinfo{person}{Emanuel Moss}, \bibinfo{person}{Elizabeth~Anne Watkins}, \bibinfo{person}{Ranjit Singh}, {and} \bibinfo{person}{Madeleine~Clare Elish}.} \bibinfo{year}{2021}\natexlab{}.
\newblock \showarticletitle{Algorithmic Impact Assessments and Accountability: The Co-Construction of Impacts}. In \bibinfo{booktitle}{\emph{Proceedings of the 2021 ACM Conference on Fairness, Accountability, and Transparency}} (Virtual Event, Canada) \emph{(\bibinfo{series}{FAccT '21})}. \bibinfo{publisher}{Association for Computing Machinery}, \bibinfo{address}{New York, NY, USA}, \bibinfo{pages}{735–746}.
\newblock
\showISBNx{9781450383097}
\urldef\tempurl%
\url{https://doi.org/10.1145/3442188.3445935}
\showURL{%
\tempurl}


\bibitem[Mitchell et~al\mbox{.}(2019)]%
        {mitchell2019model}
\bibfield{author}{\bibinfo{person}{Margaret Mitchell}, \bibinfo{person}{Simone Wu}, \bibinfo{person}{Andrew Zaldivar}, \bibinfo{person}{Parker Barnes}, \bibinfo{person}{Lucy Vasserman}, \bibinfo{person}{Ben Hutchinson}, \bibinfo{person}{Elena Spitzer}, \bibinfo{person}{Inioluwa~Deborah Raji}, {and} \bibinfo{person}{Timnit Gebru}.} \bibinfo{year}{2019}\natexlab{}.
\newblock \showarticletitle{Model cards for model reporting}. In \bibinfo{booktitle}{\emph{Proceedings of the conference on fairness, accountability, and transparency}}. \bibinfo{pages}{220--229}.
\newblock


\bibitem[Narkar et~al\mbox{.}(2021)]%
        {narkar2021model}
\bibfield{author}{\bibinfo{person}{Shweta Narkar}, \bibinfo{person}{Yunfeng Zhang}, \bibinfo{person}{Q~Vera Liao}, \bibinfo{person}{Dakuo Wang}, {and} \bibinfo{person}{Justin~D Weisz}.} \bibinfo{year}{2021}\natexlab{}.
\newblock \showarticletitle{Model LineUpper: Supporting Interactive Model Comparison at Multiple Levels for AutoML}. In \bibinfo{booktitle}{\emph{26th International Conference on Intelligent User Interfaces}}. \bibinfo{pages}{170--174}.
\newblock


\bibitem[Nilsson et~al\mbox{.}(2016)]%
        {nilsson2016applying}
\bibfield{author}{\bibinfo{person}{Tommy Nilsson}, \bibinfo{person}{Carl Hogsden}, \bibinfo{person}{Charith Perera}, \bibinfo{person}{Saeed Aghaee}, \bibinfo{person}{David~J Scruton}, \bibinfo{person}{Andreas Lund}, {and} \bibinfo{person}{Alan~F Blackwell}.} \bibinfo{year}{2016}\natexlab{}.
\newblock \showarticletitle{Applying seamful design in location-based mobile museum applications}.
\newblock \bibinfo{journal}{\emph{ACM Transactions on Multimedia Computing, Communications, and Applications (TOMM)}} \bibinfo{volume}{12}, \bibinfo{number}{4} (\bibinfo{year}{2016}), \bibinfo{pages}{1--23}.
\newblock


\bibitem[Passi(2021)]%
        {samirthesis}
\bibfield{author}{\bibinfo{person}{Samir Passi}.} \bibinfo{year}{2021}\natexlab{}.
\newblock \emph{\bibinfo{title}{Making Data Work: The Human and Organizational Lifeworlds of Data Science Practices}}.
\newblock \bibinfo{thesistype}{Ph.\,D. Dissertation}.
\newblock
\showISBNx{9798762181471}


\bibitem[Passi and Barocas(2019)]%
        {passibarocas2019}
\bibfield{author}{\bibinfo{person}{Samir Passi} {and} \bibinfo{person}{Solon Barocas}.} \bibinfo{year}{2019}\natexlab{}.
\newblock \showarticletitle{Problem Formulation and Fairness}. In \bibinfo{booktitle}{\emph{Proceedings of the Conference on Fairness, Accountability, and Transparency}} (Atlanta, GA, USA) \emph{(\bibinfo{series}{FAT* '19})}. \bibinfo{publisher}{Association for Computing Machinery}, \bibinfo{address}{New York, NY, USA}, \bibinfo{pages}{39–48}.
\newblock
\showISBNx{9781450361255}
\urldef\tempurl%
\url{https://doi.org/10.1145/3287560.3287567}
\showDOI{\tempurl}


\bibitem[Passi and Jackson(2018)]%
        {Passitrust}
\bibfield{author}{\bibinfo{person}{Samir Passi} {and} \bibinfo{person}{Steven~J. Jackson}.} \bibinfo{year}{2018}\natexlab{}.
\newblock \showarticletitle{Trust in Data Science: Collaboration, Translation, and Accountability in Corporate Data Science Projects}.
\newblock \bibinfo{journal}{\emph{Proc. ACM Hum.-Comput. Interact.}} \bibinfo{volume}{2}, \bibinfo{number}{CSCW}, Article \bibinfo{articleno}{136} (\bibinfo{date}{nov} \bibinfo{year}{2018}), \bibinfo{numpages}{28}~pages.
\newblock
\urldef\tempurl%
\url{https://doi.org/10.1145/3274405}
\showDOI{\tempurl}


\bibitem[Passi and Sengers(2020)]%
        {Passimaking}
\bibfield{author}{\bibinfo{person}{Samir Passi} {and} \bibinfo{person}{Phoebe Sengers}.} \bibinfo{year}{2020}\natexlab{}.
\newblock \showarticletitle{Making data science systems work}.
\newblock \bibinfo{journal}{\emph{Big Data \& Society}} \bibinfo{volume}{7}, \bibinfo{number}{2} (\bibinfo{year}{2020}), \bibinfo{pages}{2053951720939605}.
\newblock
\urldef\tempurl%
\url{https://doi.org/10.1177/2053951720939605}
\showDOI{\tempurl}


\bibitem[Passi and Vorvoreanu(2022)]%
        {passi2022overreliance}
\bibfield{author}{\bibinfo{person}{Samir Passi} {and} \bibinfo{person}{Mihaela Vorvoreanu}.} \bibinfo{year}{2022}\natexlab{}.
\newblock \bibinfo{booktitle}{\emph{Overreliance on AI: Literature Review}}.
\newblock \bibinfo{type}{{T}echnical {R}eport} MSR-TR-2022-12. \bibinfo{institution}{Microsoft}.
\newblock
\urldef\tempurl%
\url{https://www.microsoft.com/en-us/research/publication/overreliance-on-ai-literature-review/}
\showURL{%
\tempurl}


\bibitem[Peters et~al\mbox{.}(2020)]%
        {PetersRAI}
\bibfield{author}{\bibinfo{person}{Dorian Peters}, \bibinfo{person}{Karina Vold}, \bibinfo{person}{Diana Robinson}, {and} \bibinfo{person}{Rafael~A. Calvo}.} \bibinfo{year}{2020}\natexlab{}.
\newblock \showarticletitle{Responsible AI—Two Frameworks for Ethical Design Practice}.
\newblock \bibinfo{journal}{\emph{IEEE Transactions on Technology and Society}} \bibinfo{volume}{1}, \bibinfo{number}{1} (\bibinfo{year}{2020}), \bibinfo{pages}{34--47}.
\newblock
\urldef\tempurl%
\url{https://doi.org/10.1109/TTS.2020.2974991}
\showDOI{\tempurl}


\bibitem[Poursabzi-Sangdeh et~al\mbox{.}(2021)]%
        {poursabzi2021manipulating}
\bibfield{author}{\bibinfo{person}{Forough Poursabzi-Sangdeh}, \bibinfo{person}{Daniel~G Goldstein}, \bibinfo{person}{Jake~M Hofman}, \bibinfo{person}{Jennifer~Wortman Wortman~Vaughan}, {and} \bibinfo{person}{Hanna Wallach}.} \bibinfo{year}{2021}\natexlab{}.
\newblock \showarticletitle{Manipulating and measuring model interpretability}. In \bibinfo{booktitle}{\emph{Proceedings of the 2021 CHI Conference on Human Factors in Computing Systems}}. \bibinfo{pages}{1--52}.
\newblock


\bibitem[Radjou et~al\mbox{.}(2012)]%
        {radjou2012jugaad}
\bibfield{author}{\bibinfo{person}{Navi Radjou}, \bibinfo{person}{Jaideep Prabhu}, {and} \bibinfo{person}{Simone Ahuja}.} \bibinfo{year}{2012}\natexlab{}.
\newblock \bibinfo{booktitle}{\emph{Jugaad innovation: Think frugal, be flexible, generate breakthrough growth}}.
\newblock \bibinfo{publisher}{John Wiley \& Sons}.
\newblock


\bibitem[Raji et~al\mbox{.}(2020)]%
        {raji2020closing}
\bibfield{author}{\bibinfo{person}{Inioluwa~Deborah Raji}, \bibinfo{person}{Andrew Smart}, \bibinfo{person}{Rebecca~N White}, \bibinfo{person}{Margaret Mitchell}, \bibinfo{person}{Timnit Gebru}, \bibinfo{person}{Ben Hutchinson}, \bibinfo{person}{Jamila Smith-Loud}, \bibinfo{person}{Daniel Theron}, {and} \bibinfo{person}{Parker Barnes}.} \bibinfo{year}{2020}\natexlab{}.
\newblock \showarticletitle{Closing the AI accountability gap: Defining an end-to-end framework for internal algorithmic auditing}. In \bibinfo{booktitle}{\emph{Proceedings of the 2020 conference on fairness, accountability, and transparency}}. \bibinfo{pages}{33--44}.
\newblock


\bibitem[Rakova et~al\mbox{.}(2021a)]%
        {Rakova2021}
\bibfield{author}{\bibinfo{person}{Bogdana Rakova}, \bibinfo{person}{Jingying Yang}, \bibinfo{person}{Henriette Cramer}, {and} \bibinfo{person}{Rumman Chowdhury}.} \bibinfo{year}{2021}\natexlab{a}.
\newblock \showarticletitle{Where Responsible AI Meets Reality: Practitioner Perspectives on Enablers for Shifting Organizational Practices}.
\newblock \bibinfo{journal}{\emph{Proc. ACM Hum.-Comput. Interact.}} \bibinfo{volume}{5}, \bibinfo{number}{CSCW1}, Article \bibinfo{articleno}{7} (\bibinfo{date}{apr} \bibinfo{year}{2021}), \bibinfo{numpages}{23}~pages.
\newblock
\urldef\tempurl%
\url{https://doi.org/10.1145/3449081}
\showDOI{\tempurl}


\bibitem[Rakova et~al\mbox{.}(2021b)]%
        {rakova2021responsible}
\bibfield{author}{\bibinfo{person}{Bogdana Rakova}, \bibinfo{person}{Jingying Yang}, \bibinfo{person}{Henriette Cramer}, {and} \bibinfo{person}{Rumman Chowdhury}.} \bibinfo{year}{2021}\natexlab{b}.
\newblock \showarticletitle{Where responsible AI meets reality: Practitioner perspectives on enablers for shifting organizational practices}.
\newblock \bibinfo{journal}{\emph{Proceedings of the ACM on Human-Computer Interaction}} \bibinfo{volume}{5}, \bibinfo{number}{CSCW1} (\bibinfo{year}{2021}), \bibinfo{pages}{1--23}.
\newblock


\bibitem[Ravoka(2022)]%
        {ravoka_2022}
\bibfield{author}{\bibinfo{person}{Bogdana Ravoka}.} \bibinfo{year}{2022}\natexlab{}.
\newblock \showarticletitle{Slowing Down AI with Speculative Friction}.
\newblock \bibinfo{journal}{\emph{Branch Magazine}} (\bibinfo{year}{2022}).
\newblock


\bibitem[Rosson and Carroll(2009)]%
        {rosson2009scenario}
\bibfield{author}{\bibinfo{person}{Mary~Beth Rosson} {and} \bibinfo{person}{John~M Carroll}.} \bibinfo{year}{2009}\natexlab{}.
\newblock \showarticletitle{Scenario based design}.
\newblock \bibinfo{journal}{\emph{Human-computer interaction. boca raton, FL}} (\bibinfo{year}{2009}), \bibinfo{pages}{145--162}.
\newblock


\bibitem[Rubambiza et~al\mbox{.}(2022)]%
        {phoebeseamless2022}
\bibfield{author}{\bibinfo{person}{Gloire Rubambiza}, \bibinfo{person}{Phoebe Sengers}, {and} \bibinfo{person}{Hakim Weatherspoon}.} \bibinfo{year}{2022}\natexlab{}.
\newblock \showarticletitle{Seamless Visions, Seamful Realities: Anticipating Rural Infrastructural Fragility in Early Design of Digital Agriculture}. In \bibinfo{booktitle}{\emph{Proceedings of the 2022 CHI Conference on Human Factors in Computing Systems}} (New Orleans, LA, USA) \emph{(\bibinfo{series}{CHI '22})}. \bibinfo{publisher}{Association for Computing Machinery}, \bibinfo{address}{New York, NY, USA}, Article \bibinfo{articleno}{451}, \bibinfo{numpages}{15}~pages.
\newblock
\showISBNx{9781450391573}
\urldef\tempurl%
\url{https://doi.org/10.1145/3491102.3517579}
\showDOI{\tempurl}


\bibitem[Selbst et~al\mbox{.}(2019)]%
        {selbst_fairness_2019}
\bibfield{author}{\bibinfo{person}{Andrew~D. Selbst}, \bibinfo{person}{Danah Boyd}, \bibinfo{person}{Sorelle~A. Friedler}, \bibinfo{person}{Suresh Venkatasubramanian}, {and} \bibinfo{person}{Janet Vertesi}.} \bibinfo{year}{2019}\natexlab{}.
\newblock \showarticletitle{Fairness and {Abstraction} in {Sociotechnical} {Systems}}. In \bibinfo{booktitle}{\emph{Proceedings of the {Conference} on {Fairness}, {Accountability}, and {Transparency}}}. \bibinfo{publisher}{ACM}, \bibinfo{address}{Atlanta GA USA}, \bibinfo{pages}{59--68}.
\newblock
\showISBNx{978-1-4503-6125-5}
\urldef\tempurl%
\url{https://doi.org/10.1145/3287560.3287598}
\showDOI{\tempurl}


\bibitem[Shelby et~al\mbox{.}(2023)]%
        {ShelbySociotechnicalHarms}
\bibfield{author}{\bibinfo{person}{Renee Shelby}, \bibinfo{person}{Shalaleh Rismani}, \bibinfo{person}{Kathryn Henne}, \bibinfo{person}{AJung Moon}, \bibinfo{person}{Negar Rostamzadeh}, \bibinfo{person}{Paul Nicholas}, \bibinfo{person}{N'Mah Yilla}, \bibinfo{person}{Jess Gallegos}, \bibinfo{person}{Andrew Smart}, \bibinfo{person}{Emilio Garcia}, {and} \bibinfo{person}{Gurleen Virk}.} \bibinfo{year}{2023}\natexlab{}.
\newblock \bibinfo{title}{Identifying Sociotechnical Harms of Algorithmic Systems: Scoping a Taxonomy for Harm Reduction}.
\newblock
\newblock
\showeprint[arxiv]{2210.05791}~[cs.HC]


\bibitem[Shneiderman(2021)]%
        {shneiderman2021responsible}
\bibfield{author}{\bibinfo{person}{Ben Shneiderman}.} \bibinfo{year}{2021}\natexlab{}.
\newblock \showarticletitle{Responsible AI: Bridging from ethics to practice}.
\newblock \bibinfo{journal}{\emph{Commun. ACM}} \bibinfo{volume}{64}, \bibinfo{number}{8} (\bibinfo{year}{2021}), \bibinfo{pages}{32--35}.
\newblock


\bibitem[Sloane et~al\mbox{.}(2020)]%
        {sloane2020participation}
\bibfield{author}{\bibinfo{person}{Mona Sloane}, \bibinfo{person}{Emanuel Moss}, \bibinfo{person}{Olaitan Awomolo}, {and} \bibinfo{person}{Laura Forlano}.} \bibinfo{year}{2020}\natexlab{}.
\newblock \showarticletitle{Participation is not a design fix for machine learning}.
\newblock \bibinfo{journal}{\emph{arXiv preprint arXiv:2007.02423}} (\bibinfo{year}{2020}).
\newblock


\bibitem[Strauss and Corbin(1990)]%
        {Strauss1990}
\bibfield{author}{\bibinfo{person}{Anselm Strauss} {and} \bibinfo{person}{Juliet~M. Corbin}.} \bibinfo{year}{1990}\natexlab{}.
\newblock \bibinfo{booktitle}{\emph{{Basics of Qualitative Research: Grounded Theory Techniques and Procedures}}}.
\newblock \bibinfo{publisher}{Sage}, \bibinfo{address}{New York}.
\newblock


\bibitem[Suresh et~al\mbox{.}(2021)]%
        {suresh2021beyond}
\bibfield{author}{\bibinfo{person}{Harini Suresh}, \bibinfo{person}{Steven~R Gomez}, \bibinfo{person}{Kevin~K Nam}, {and} \bibinfo{person}{Arvind Satyanarayan}.} \bibinfo{year}{2021}\natexlab{}.
\newblock \showarticletitle{Beyond Expertise and Roles: A Framework to Characterize the Stakeholders of Interpretable Machine Learning and their Needs}. In \bibinfo{booktitle}{\emph{Proceedings of the 2021 CHI Conference on Human Factors in Computing Systems}}. \bibinfo{pages}{1--16}.
\newblock


\bibitem[Szymanski et~al\mbox{.}(2021)]%
        {szymanski2021visual}
\bibfield{author}{\bibinfo{person}{Maxwell Szymanski}, \bibinfo{person}{Martijn Millecamp}, {and} \bibinfo{person}{Katrien Verbert}.} \bibinfo{year}{2021}\natexlab{}.
\newblock \showarticletitle{Visual, textual or hybrid: the effect of user expertise on different explanations}. In \bibinfo{booktitle}{\emph{26th International Conference on Intelligent User Interfaces}}. \bibinfo{pages}{109--119}.
\newblock


\bibitem[Thurnes et~al\mbox{.}(2015)]%
        {thurnes2015using}
\bibfield{author}{\bibinfo{person}{Christian~M Thurnes}, \bibinfo{person}{Frank Zeihsel}, \bibinfo{person}{Svetlana Visnepolschi}, {and} \bibinfo{person}{Frank Hallfell}.} \bibinfo{year}{2015}\natexlab{}.
\newblock \showarticletitle{Using TRIZ to invent failures--concept and application to go beyond traditional FMEA}.
\newblock \bibinfo{journal}{\emph{Procedia engineering}}  \bibinfo{volume}{131} (\bibinfo{year}{2015}), \bibinfo{pages}{426--450}.
\newblock


\bibitem[VanGundy(1984)]%
        {vangundy1984brain}
\bibfield{author}{\bibinfo{person}{Arthur~B VanGundy}.} \bibinfo{year}{1984}\natexlab{}.
\newblock \showarticletitle{Brain writing for new product ideas: an alternative to brainstorming}.
\newblock \bibinfo{journal}{\emph{Journal of Consumer Marketing}} (\bibinfo{year}{1984}).
\newblock


\bibitem[Vertesi(2014)]%
        {vertesi2014seamful}
\bibfield{author}{\bibinfo{person}{Janet Vertesi}.} \bibinfo{year}{2014}\natexlab{}.
\newblock \showarticletitle{Seamful spaces: Heterogeneous infrastructures in interaction}.
\newblock \bibinfo{journal}{\emph{Science, Technology, \& Human Values}} \bibinfo{volume}{39}, \bibinfo{number}{2} (\bibinfo{year}{2014}), \bibinfo{pages}{264--284}.
\newblock


\bibitem[Vorvoreanu et~al\mbox{.}(2023)]%
        {vorvoreanu2023responsible}
\bibfield{author}{\bibinfo{person}{Mihaela Vorvoreanu}, \bibinfo{person}{Amy Heger}, \bibinfo{person}{Samir Passi}, \bibinfo{person}{Shipi Dhanorkar}, \bibinfo{person}{Zoe Kahn}, {and} \bibinfo{person}{Ruotong Wang}.} \bibinfo{year}{2023}\natexlab{}.
\newblock \bibinfo{booktitle}{\emph{Responsible AI Maturity Model}}.
\newblock \bibinfo{type}{{T}echnical {R}eport} MSR-TR-2023-26. \bibinfo{institution}{Microsoft}.
\newblock
\urldef\tempurl%
\url{https://www.microsoft.com/en-us/research/publication/responsible-ai-maturity-model/}
\showURL{%
\tempurl}


\bibitem[Wang et~al\mbox{.}(2019)]%
        {wang2019designing}
\bibfield{author}{\bibinfo{person}{Danding Wang}, \bibinfo{person}{Qian Yang}, \bibinfo{person}{Ashraf Abdul}, {and} \bibinfo{person}{Brian~Y Lim}.} \bibinfo{year}{2019}\natexlab{}.
\newblock \showarticletitle{Designing theory-driven user-centric explainable AI}. In \bibinfo{booktitle}{\emph{Proceedings of the 2019 CHI conference on human factors in computing systems}}. \bibinfo{pages}{1--15}.
\newblock


\bibitem[Weiser(1994)]%
        {weiser1994creating}
\bibfield{author}{\bibinfo{person}{Mark Weiser}.} \bibinfo{year}{1994}\natexlab{}.
\newblock \showarticletitle{Creating the invisible interface: (invited talk)}. In \bibinfo{booktitle}{\emph{Proceedings of the 7th annual ACM symposium on User interface software and technology}}. \bibinfo{pages}{1}.
\newblock


\bibitem[Xie et~al\mbox{.}(2020)]%
        {xie2020chexplain}
\bibfield{author}{\bibinfo{person}{Yao Xie}, \bibinfo{person}{Melody Chen}, \bibinfo{person}{David Kao}, \bibinfo{person}{Ge Gao}, {and} \bibinfo{person}{Xiang'Anthony' Chen}.} \bibinfo{year}{2020}\natexlab{}.
\newblock \showarticletitle{CheXplain: enabling physicians to explore and understand data-driven, AI-enabled medical imaging analysis}. In \bibinfo{booktitle}{\emph{Proceedings of the 2020 CHI Conference on Human Factors in Computing Systems}}. \bibinfo{pages}{1--13}.
\newblock


\bibitem[Yang et~al\mbox{.}(2020)]%
        {yang2020re}
\bibfield{author}{\bibinfo{person}{Qian Yang}, \bibinfo{person}{Aaron Steinfeld}, \bibinfo{person}{Carolyn Ros{\'e}}, {and} \bibinfo{person}{John Zimmerman}.} \bibinfo{year}{2020}\natexlab{}.
\newblock \showarticletitle{Re-examining whether, why, and how human-AI interaction is uniquely difficult to design}. In \bibinfo{booktitle}{\emph{Proceedings of the 2020 chi conference on human factors in computing systems}}. \bibinfo{pages}{1--13}.
\newblock


\bibitem[Zhang et~al\mbox{.}(2020)]%
        {zhang2020effect}
\bibfield{author}{\bibinfo{person}{Yunfeng Zhang}, \bibinfo{person}{Q~Vera Liao}, {and} \bibinfo{person}{Rachel~KE Bellamy}.} \bibinfo{year}{2020}\natexlab{}.
\newblock \showarticletitle{Effect of confidence and explanation on accuracy and trust calibration in AI-assisted decision making}. In \bibinfo{booktitle}{\emph{Proceedings of the 2020 Conference on Fairness, Accountability, and Transparency}}. \bibinfo{pages}{295--305}.
\newblock


\end{thebibliography}

% appendix
\newpage
\appendix

\section{Appendix}
\subsection{High Resolution Picture of the Whiteboard} \label{sec:appendix_high_res_whiteboard}

\begin{figure}[h]
    \centering
    \includegraphics[width=\textwidth]{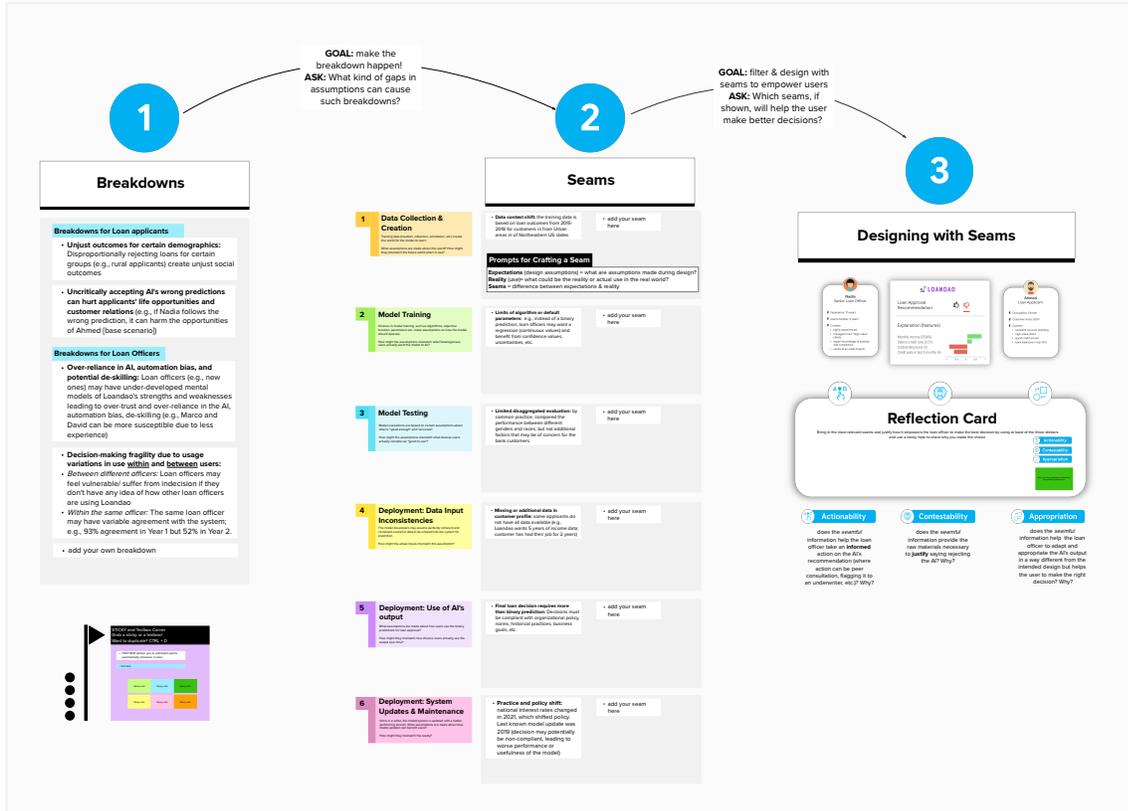}
    \caption{A higher resolution picture of Figure~\ref{fig:mural_design_process} showing a screenshot of the virtual whiteboard used for the seamful XAI design activity in the study, with zoomed-in examples. \textbf{Area 1:} Envisioning breakdown (Step 1). In the study, we provided sample breakdowns, which participants could either use directly or get inspiration for their own envisioning. \textbf{Area 2:} Anticipating \& crafting seams (Step 2). We provided guiding prompts for crafting the seams can procedurally guide the process of writing down the seam in an effective manner. We also shared exemplary seams for each stage of the AI lifecycle framework. \textbf{Area 3:} Designing with seams. We asked participants to articulate their reasoning for choosing a seam and tag which user goals the selected seam can support for augmenting user agency. If viewed on a PDF viewer like Adobe Acrobat, the reader can zoom in as necessary to read the text on this board.}
    \Description[figure]{A screenshot of the virtual whiteboard used for the Seamful XAI design activity in the study, with zoomed-in examples. There are three main areas highlighted. Area 1: Envisioning breakdowns (this is step 1). In this study, we provided sample breakdowns (highlighted in the A box - e.g., decision making fragility due to usage variations between users). Participants could either use these directly or inspire their own envisioning. Area 2: Anticipating and crafting seams (this is step 2). We provided guiding prompts (shown in B box - e.g., Goal is to make the breakdown happen and the ask is to identify which kinds of gaps in assumptions can cause such breakdowns). We also provided exemplary seams (shown in box C - e.g., shifting practices and policies).  Area 3: Designing with seams (this is step 3). We asked participants to articulate and tag which user goals the selected seam (shown in box E - e.g., actionability, contestability, or appropriation) can support for augmenting user agency. The virtual whiteboard is organized as a tripartite table with a combination of text and icons.}
    \label{fig:highres_whiteboard}
\end{figure}

\newpage
\subsection{High Resolution Picture of the Bird's-eye View of All Seams} \label{sec:appendix_high_res_allseams}

\begin{figure}[h]
    \centering
    \includegraphics[width=\textwidth]{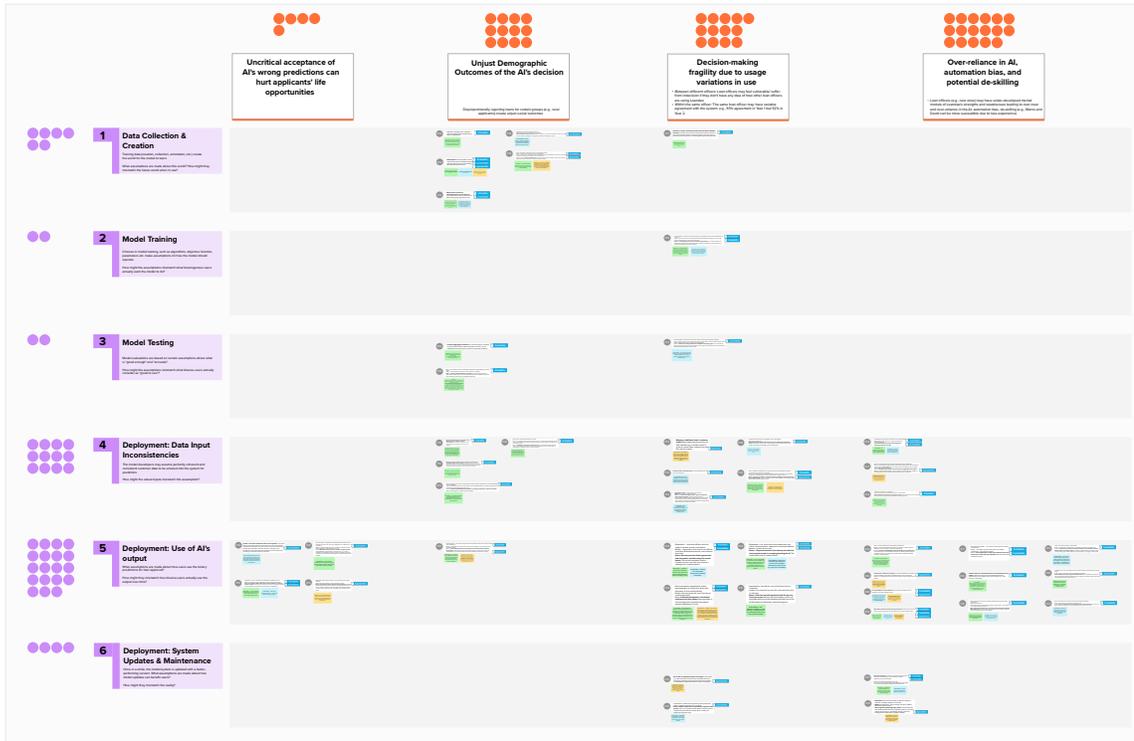}
    \caption{A higher resolution picture of Figure~\ref{fig:all_seams} showing a bird's eye view of all seams from all participants for all breakdowns along all AI lifecycle. Breakdowns are the columns while the AI lifecycle stages are on the rows. Te dots above breakdowns provide the number of times a particular one was. All the seams appearing below this column were crafted in connection to that particular breakdown. The dots next to each lifecycle stage showcase the number of seams crafted for that stage. Each row has the seams along with the justifications for how the seams enhances agency in the sticky notes. If viewed on a PDF viewer like Adobe Acrobat, the reader can zoom in as necessary to read the text on this board. 
    }
    \Description[figure]{A bird's eye view of all seams from all participants for all breakdowns along all AI lifecycle. This is in the form of a table. There are six rows signifying the different stages in the AI lifecycle: data collection and curation, model training, model testing, data input and inconsistencies, use of AI's output, and system updates and maintenance. There are dots next to the row headers that showcase the number of seams crafted for a given AI lifecycle stage (e.g., 6 seams were crafted for the data collection and curation stage). There are four columns indicating the different breakdowns identified by participants (e.g., box A highlights one such breakdown - that of overreliance on AI that was chosen for all the seams appearing below in this column. There are dots above the column header that provide the number of times this was chosen (e.g., overreliance on AI was chosen 17 times). Box C shows a zoomed-in view to provide a sample of seams crafted by our participants along with the justification for how the seams enhance agency.}
    \label{fig:highres_allseams}
\end{figure}

\end{document}